\documentclass[pra,a4paper,twocolumn,showpacs,floatfix,superscriptaddress,amsmath,preprintnumbers,citeautoscript,aps,longbibliography]{revtex4-1}
\pdfoutput=1
\usepackage[T1]{fontenc}\usepackage[latin1]{inputenc}
\usepackage{dcolumn,graphicx,color,booktabs,microtype,afterpage} \graphicspath{{Figures/}}
\usepackage[charter,greekuppercase=italicized]{mathdesign}
\renewcommand{\figurename}{Fig.}
\renewcommand{\tablename}{Table}
\makeatletter\renewcommand{\fnum@figure}[1]{\figurename~\thefigure.}\makeatother
\makeatletter\renewcommand{\fnum@table}[1]{\tablename~\thetable.}\makeatother
\newcount\hh \newcount\mm
\hh=\time \divide\hh by 60
\mm=\hh \multiply\mm by 60 \mm=-\mm
\advance\mm by \time
\def\now{\number\hh:\ifnum\mm<10{}0\fi\number\mm}
\usepackage[colorlinks,plainpages=false,linkcolor=blue,urlcolor=blue,citecolor=blue,pdfpagemode=UseNone,pdfstartview=FitBH]{hyperref}

\usepackage[autoplay]{animate}

\begin{document}
\title{
Decoupled spin dynamics in the rare-earth orthoferrite YbFeO$_3$: \\
Evolution of magnetic excitations through the spin-reorientation transition.}
\author{S.~E.~Nikitin}
\affiliation{Max Planck Institute for Chemical Physics of Solids, N\"{o}thnitzer Str. 40, D-01187 Dresden, Germany}
\affiliation{Institut f\"ur Festk\"orper- und Materialphysik, Technische Universit\"at Dresden, D-01069 Dresden, Germany}
\author{L.~S.~Wu}
\affiliation{Neutron Scattering Division, Oak Ridge National Laboratory, Oak Ridge, Tennessee 37831, USA}
\author{A.~S.~Sefat}
\affiliation{Materials Science and Technology Division, Oak Ridge National Laboratory, Oak Ridge, Tennessee 37831, USA}
\author{K.~A.~Shaykhutdinov}
\affiliation{Kirensky Institute of Physics, Federal Research Center, Krasnoyarsk 660036, Russia}
\author{Z.~Lu}
\affiliation{Helmholtz-Zentrum Berlin f\"ur Materialien und Energie, D-14109 Berlin, Germany}
\author{S.~Meng}
\affiliation{Helmholtz-Zentrum Berlin f\"ur Materialien und Energie, D-14109 Berlin, Germany}
\affiliation{China Institute of Atomic Energy, Beijing 102413, People's Republic of China}
\author{E.~V.~Pomjakushina}
\affiliation{Laboratory for Multiscale Materials Experiments, Paul Scherrer Institut, CH-5232 Villigen PSI, Switzerland}
\author{K.~Conder}
\affiliation{Laboratory for Multiscale Materials Experiments, Paul Scherrer Institut, CH-5232 Villigen PSI, Switzerland}
\author{G.~Ehlers}
\affiliation{Neutron Technologies Division, Oak Ridge National Laboratory, Oak Ridge, Tennessee 37831, USA}
\author{M.~D.~Lumsden}
\affiliation{Neutron Scattering Division, Oak Ridge National Laboratory, Oak Ridge, Tennessee 37831, USA}
\author{A.~I.~Kolesnikov}
\affiliation{Neutron Scattering Division, Oak Ridge National Laboratory, Oak Ridge, Tennessee 37831, USA}
\author{S.~Barilo}
\affiliation{Institute of Solid State and Semiconductor Physics, National Academy of Sciences, 220072 Minsk, Belarus}
\author{S.~A.~Guretskii}
\affiliation{Institute of Solid State and Semiconductor Physics, National Academy of Sciences, 220072 Minsk, Belarus}
\author{D.~S.~Inosov}
\affiliation{Institut f\"ur Festk\"orper- und Materialphysik, Technische Universit\"at Dresden, D-01069 Dresden, Germany}
\author{A.~Podlesnyak}
\thanks{Corresponding author: podlesnyakaa@ornl.gov}
\affiliation{Neutron Scattering Division, Oak Ridge National Laboratory, Oak Ridge, Tennessee 37831, USA}

\begin{abstract}
In this paper we present a comprehensive study of magnetic dynamics in the rare-earth orthoferrite YbFeO$_3$ at temperatures below and above the spin-reorientation (SR) transition $T_{\mathrm{SR}}=7.6$~K, in magnetic fields applied along the $a, b$ and $c$ axes. Using single-crystal inelastic neutron scattering, we observed that the spectrum of magnetic excitations consists of two collective modes well separated in energy: 3D gapped magnons with a bandwidth of $\sim$60~meV, associated with the antiferromagnetically (AFM) ordered Fe subsystem, and quasi-1D AFM fluctuations of $\sim$1~meV within the Yb subsystem, with no hybridization of those modes.
The spin dynamics of the Fe subsystem changes very little through the SR transition and could be well described in the frame of semiclassical linear spin-wave theory. On the other hand, the rotation of the net moment of the Fe subsystem at $T_{\mathrm{SR}}$ drastically changes the excitation spectrum of the Yb subsystem, inducing the transition between two regimes with magnon and spinon-like fluctuations.
At $T~<~T_{\mathrm{SR}}$, the Yb spin chains have a well defined field-induced ferromagnetic (FM) ground state, and the spectrum consists of a sharp single-magnon mode, a two-magnon bound state, and a two-magnon continuum, whereas at $T~>~T_{\mathrm{SR}}$ only a gapped broad spinon-like continuum dominates the spectrum.
In this work we show that a weak quasi-1D coupling within the Yb subsystem $J_\text{Yb-Yb}$, mainly neglected in previous studies, creates unusual quantum spin dynamics on the low energy scales.
The results of our work may stimulate further experimental search for similar compounds with several magnetic subsystems and energy scales, where low-energy fluctuations and underlying physics could be ``hidden'' by a dominating interaction.
\end{abstract}
\maketitle

\section{Introduction}

Quantum phase transitions have been a matter of special interest in condensed matter physics during the last decades~\cite{Sachdev, StockertSteglich11, SiSteglich10}. In contrast to the classical phase transitions induced by thermal fluctuations, quantum phase transitions are driven by quantum fluctuations and can be induced by an external tuning parameter, like pressure, magnetic field, uniaxial strain etc. Among all quantum critical systems, the antiferromagnetic (AFM) Heisenberg $S=\frac{1}{2}$ chain is one of the  simplest examples: at zero field it has a tangled singlet ground state and fractionalized magnetic excitations, so-called ``spinons'' carrying spin $\frac{1}{2}$~\cite{haldane1991spinon, tennant1993unbound}, whereas in a magnetic field it undergoes a transition into the field-polarized state, with a well defined classical FM ground state and $S=1$ magnon quasiparticles as elementary excitations~\cite{Mourigal}.
In this work we studied the spin dynamics of YbFeO$_3$, which contains two magnetic sublattices and observed an intriguing coexistence of the classical high-energy spin waves and unconventional low-energy spin excitations, which spontaneously transform from classical magnon to quantum spinon quasiparticles with increasing temperature.

\begin{figure*}[tb]
    \includegraphics[width=2\columnwidth]{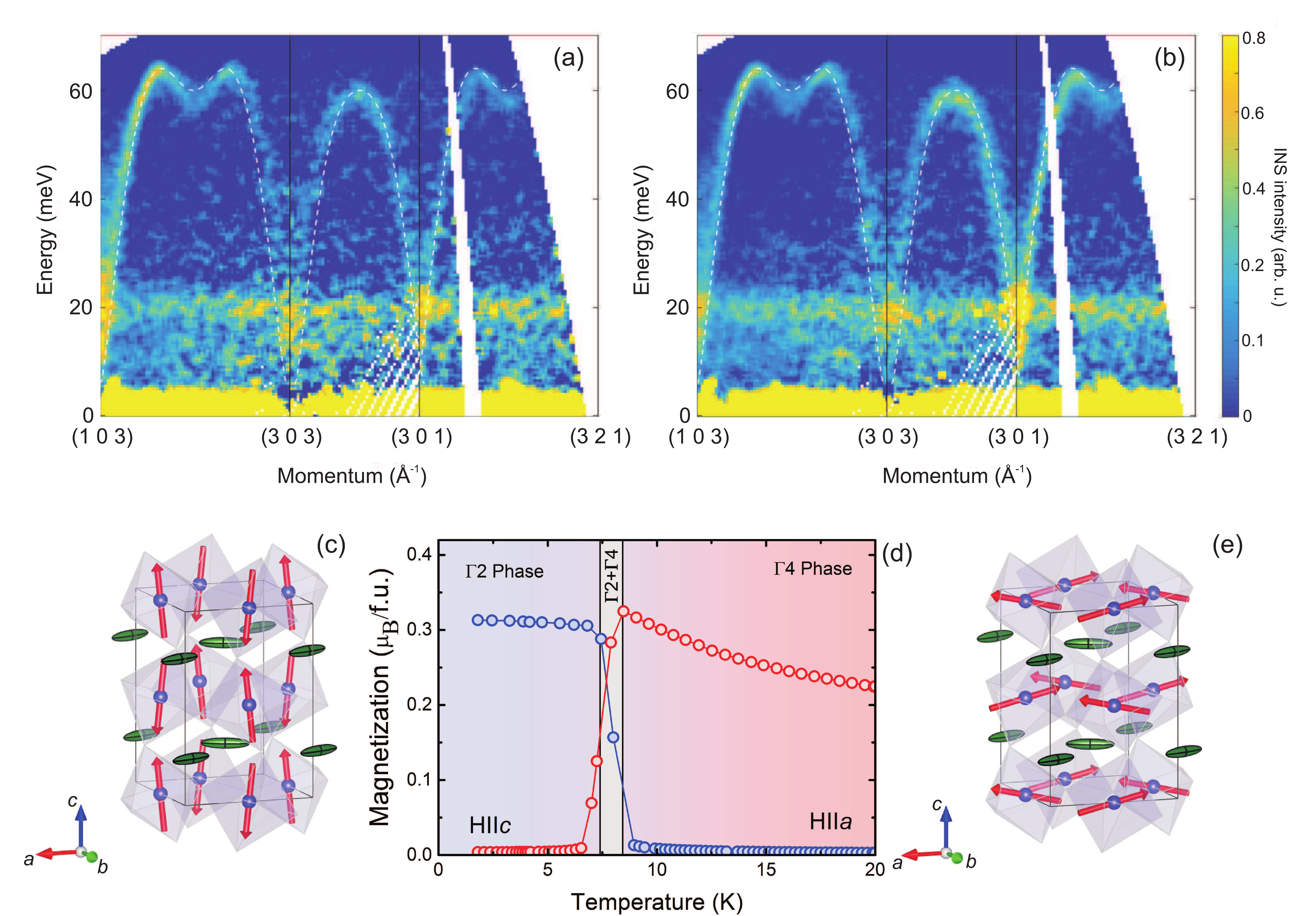}
    \caption{~(a,b) Magnon excitations in YbFeO$_3$ along (103)-(303)-(301)-(321) directions of reciprocal space taken at temperatures below ($T=2$~K, left) and above ($T=15$~K, right) the SR transition (indices are given in reciprocal lattice units). Dotted lines represent results of the linear spin-wave calculations.
    (c,e) Sketches of magnetic structures of YbFeO$_3$ below (c) and above (e) the SR transition. Blue spheres show Fe ions, green ellipsoids represent anisotropic magnetic moments of Yb. In the magnetic phase $\rm \Gamma2$ (c), below $T_{\mathrm{SR}}$, Fe moments align along the $c$ axis, and spin canting results in a net moment along the $a$ axis. Above $T_{\mathrm{SR}}$ ($\rm \Gamma4$ phase), Fe moments rotate to the $a$ axis, and spin canting gives a net moment along the $c$ axis.
    (d) Temperature dependences of the magnetization of YbFeO$_3$ measured at $H=0.01$~T along the $c$ (red) and $a$ (blue) axes.}
    \label{SEQUOIA_slices}
\end{figure*}

YbFeO$_3$ belongs to the family of iron-based orthorhombic perovskites, $R$FeO$_3$ ($R$ -- rare-earth, Bi or Y), which attract considerable attention due to the high-temperature multiferroic properties of BiFeO$_3$~\cite{Cheong,Khomskii}, anisotropic magnetic entropy evolution~\cite{ke2016anisotropic}, laser-pulse induced ultrafast spin-reorientation~\cite{Kimel, de2011laser, jiang2013dynamical} etc. Magnetic property investigations of the rare-earth orthoferrites $R$FeO$_3$ have shown that the Fe$^{3+}$ moments ($S=\frac{5}{2}$) are ordered in a canted AFM structure $\rm \Gamma4$ at high temperature with $T_{\rm N}\approx600$~K (details of the notations are given in~\cite{White}), and the spin canting gives a weak net ferromagnetic moment along the $c$ axis [Fig.~\ref{SEQUOIA_slices}(c)]~\cite{bozorth1958magnetization, White, bazaliy2005measurements}. Furthermore, symmetry analysis and careful neutron diffraction measurements have found a second ``hidden'' canting along the $b$-axis, which is symmetric relative to the $ac$-plane and does not create a net moment~\cite{plakhtij1981experimental, plakhty1983neutron}.
With decreasing temperature, a spontaneous spin-reorientation (SR) transition from $\rm \Gamma4$ to the $\rm \Gamma2$ magnetic configuration occurs in many orthoferrites with magnetic $R$-ions~\cite{White, bozorth1958magnetization} in a wide temperature range from $T_{\mathrm{SR}}\approx450$~K for SmFeO$_3$ down to $T_{\mathrm{SR}}\approx7.6$~K for YbFeO$_3$, and the net magnetic moment rotates from the $a$ to the $c$ axis [see Fig.~\ref{SEQUOIA_slices}(c-e)]. Most of previous work that was devoted to the investigation of the SR transition in $R$FeO$_3$, associated this phenomenon with the $R$-Fe exchange interaction, because orthoferrites with nonmagnetic $R=$La, Y or Lu preserve the $\rm \Gamma4$ magnetic structure down to the lowest temperatures.

Taking into account three characteristic temperatures: $T_{\rm N}^{\rm Fe}\sim600$~K, $T_{\mathrm{SR}}\sim10$~K and $T_{\rm N}^{\rm Yb}\sim1$~K (known for the isostructural YbAlO$_3$~\cite{radhakrishna1981antiferromagnetic}) one could expect a similar hierarchy of the exchange interactions $J_\text{Fe-Fe}\gg{}J_\text{Fe-Yb}\gg{}J_\text{Yb-Yb}$ and multiple magnetic modes, corresponding to each of the energy scales. From the experimental point of view, the best experimental technique to study the details of the magnetic interaction is the inelastic neutron scattering. However, to the best of our knowledge, investigations of the spin dynamics in the orthoferrites were mainly focused on the Fe subsystem. Results of the INS experiments have shown, that the Fe spin fluctuations are dominated by the high-energy gapped magnons with an energy scale of $E\approx60$~meV and could be reasonably well described using a simple linear spin-wave theory (LSWT)~\cite{Hahn, park2017low, shapiro1974, gukasov1997}, while the details regarding the dispersion of magnetic modes, associated with $R$-Fe and $R$-$R$ exchange interactions, were mainly unexplored.

In this paper we present the results of a detailed study of the spin dynamics in YbFeO$_3$ that covers the energy scales mentioned above. We observed the high-energy spin-wave modes within the Fe-subsystem at $E\approx4-65$~meV, which are almost unaffected by the SR transition. Well below the gap of the Fe excitations $\Delta\approx4$~meV, we observed a second gapped excitation, with dispersion along the $c$ axis only, which can be associated with the fluctuations of the Yb moments coupled in quasi-1D XXZ spin chains. The most remarkable outcome of our work is an unusual low-dimensional spin dynamics of the highly anisotropic Yb subsystem, which significantly changes through the SR transition. Below $T_{\mathrm{SR}}$, Yb moments are fully polarized by the effective Fe field, giving rise to the conventional magnons accompanied by a higher-energy 2-magnon bound state and a broad continuum. On the other hand, above $T_{\mathrm{SR}}$, an effective field is transverse to the easy axis, leading to the nonpolarized ground state and to the rise of unconventional spinon excitations, which are clearly seen as a broad continuum above the single-particle mode in the excitation spectrum. INS measurements of low-energy spin dynamics under magnetic field along different axes show that the external magnetic field has a similar effect as the effective internal field, induced by the Fe subsystem.

\section{Experimental details}

INS experiments were carried out on two YbFeO$_3$ single crystals with the masses of $\sim$3.8~g (used in time-of-flight (TOF) measurements on the SEQUOIA and CNCS instruments) and $\sim$1.2~g (for measurements on the triple-axis spectrometer (TAS) FLEXX) with a mosaicity $\leqslant$1$^{\circ}$. The crystals were grown by the floating-zone method and using the fluxed melt  crystallization (on seeds) technique, respectively (see~\cite{SI} for details).
Most of the INS measurements were performed using TOF spectrometers: Cold Neutron Chopper Spectrometer (CNCS)~\cite{CNCS1,CNCS2} and SEQUOIA~\cite{granroth2010sequoia} at the Spallation Neutron Source (SNS) at Oak Ridge National Laboratory. For the high-energy measurements on the SEQUOIA instrument we fixed the incident neutron energy $E_{\rm i}=100$~meV and oriented the sample with the $[010]$ direction vertically. Data were taken at temperatures above ($T=15$~K) and below ($T=2$~K) the SR transition.
For the low-energy measurements we used the CNCS instrument. The sample was measured in two orientations, with either $[100]$ or $[010]$ directions pointed vertically, and the vertical magnetic field was applied along the $a$ and $b$ axes, respectively. The measurements were carried out using the rotating single crystal method at temperatures of $T=$~2~K and 10~K. The data were collected using a fixed incident neutron energy of $E_{\rm i}=3.0$~meV resulting in a full-width at half-maximum energy resolution of 0.07~meV at the elastic position.

All time-of-flight datasets were combined to produce a four-dimensional scattering-intensity function $I(\mathbf{Q},\hbar\omega)$, where $\mathbf{Q}$ is the momentum transfer and $\hbar\omega$ is the energy transfer. For data reduction and analysis we used the \textsc{Mantid}~\cite{Mantid}, \textsc{Horace}~\cite{Horace} and \textsc{SpinW}~\cite{Toth} software packages. For the crystal electric field (CEF) calculations and numerical diagonalization of the 1D-XXZ Hamiltonian, we used \textsc{McPhase}~\cite{McPhase} and \textsc{ALPS}~\cite{ALPS1, ALPS2} software, respectively.

Low-energy INS measurements with horizontal magnetic field applied along the $c$ axis were performed using the cold-neutron triple-axis spectrometer FLEXX~(V2)~\cite{le2013gains} with the HM-1 magnet at the Helmholtz-Zentrum Berlin (HZB). The sample was mounted and mechanically fixed in a special aluminum container in order to avoid magnetic field-induced torque due to a strong anisotropy of magnetization of YbFeO$_3$ at low temperatures. Measurements were carried out with a fixed final energy $(k_{\mathrm{f}}=1.3$~\AA$^{-1}$) at temperatures between 2 and 10~K and magnetic fields up to $H=4$~T.

Specific-heat measurements were carried out using a commercial PPMS-6000 from Quantum Design in magnetic fields up to 12~T applied along the $a$-axis. Magnetization curves were measured using vibrating-sample magnetometer MPMS-3 with magnetic field up to 7~T applied along the $a$ and $c$ axes.

 \begin{figure}[tb]
    \includegraphics[width=1\columnwidth]{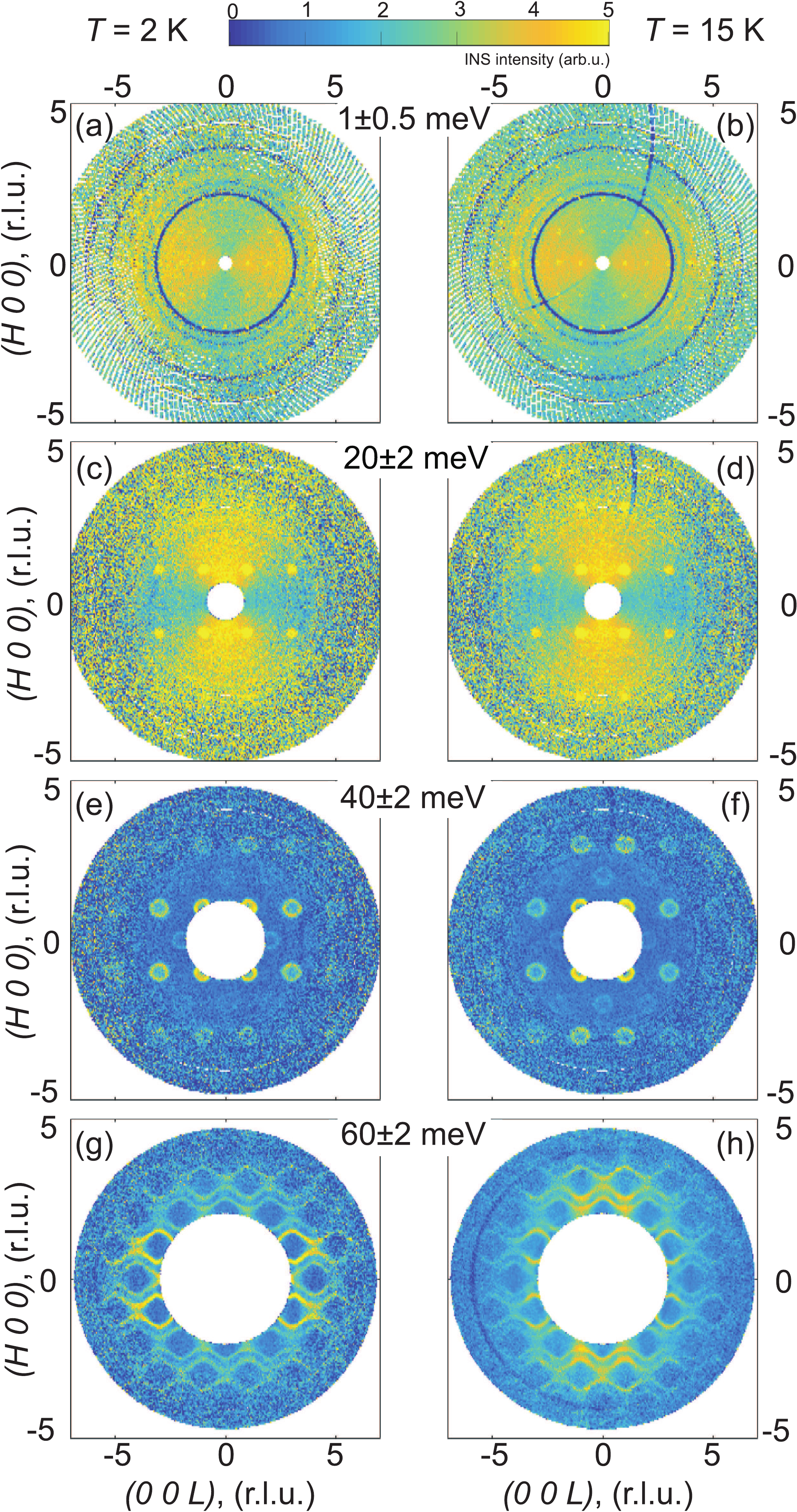}
    \caption{~Constant energy slices of the INS intensity wthin the $(H0L)$ scattering plane taken at $T=2$~K (a,c,e,g) and $T=15$~K (b,d,f,h). The scattering intensities were integrated within energy windows, indicated between the corresponding panels. The intensities of the (a) and (b) panels have been scaled $\times 0.1$ due to the proximity of the elastic line.
}
    \label{SEQUOIA_const_E}
\end{figure}

\section{Zero field measurements: experimental results}
\subsection{High-energy INS data}

The spin dynamics of the Fe subsystem of rare-earth orthoferrites with various rare-earth ions ($R=$~Lu, Y, Tm, Er) were a matter of comprehensive investigations \cite{Hahn, shapiro1974, gukasov1997, park2017low}.
However, to the best of our knowledge, the details of spin dynamics of YbFeO$_3$ have not been published yet and we start the discussion of our INS data with the report of the high-energy spin dynamics.
Figure~\ref{SEQUOIA_slices}(a,b) presents experimental INS spectra along all principal ${Q}_{{H}}$, ${Q}_{{K}}$ and ${Q}_{{L}}$ directions, taken at temperatures $T=2$ and 15~K, below and above the SR transition, respectively. Observed magnon branches stem from the magnetic Bragg peaks with an even sum of $H+K+L$, and the maximum energy of spin-wave branches $E_{\mathrm{max}}~\approx~65$~meV is similar to that observed in other orthoferrites and could be clearly associated with a collective excitation of the Fe$^{3+}$ magnetic moments.
The horizontal dispersionless line at $E~\approx~20$~meV was associated with the Yb$^{3+}$ single-ion CEF transition from the ground state to the first excited doublet~(see the CEF calculations in~\cite{SI}).

Figure~\ref{SEQUOIA_const_E} shows constant-energy slices in the $(H0L)$ plane taken around energies $E=1, 20, 40, 60$~meV at $T = 2$~K (left) and $T = 15$~K (right). Slices at $E = 40, 60$~meV show the clean spin-wave excitations caused by the Fe-Fe interaction for both temperatures, and one can see the redistribution of the INS intensity, which is concentrated either along the $L$ or $H$ direction, at $T=15$~K and 2~K, respectively, as expected from the known SR transition of the Fe moments. For $E=1$ and 20~meV one can see additional intensity, which corresponds to the ground state splitting and first excited CEF doublet of Yb$^{3+}$, respectively. In contrast to the conventional CEF excitations without significant $\mathbf{Q}$-dependence, here one can see that the INS intensity has an X-shape (hourglass) for both temperatures, which does not change through the SR transition for both excitations. 

Due to the experimental resolution limitations ($\Delta{}E\approx3$~meV for our setup) we were not able to precisely extract the gap values from the SEQUOIA datasets and performed additional measurements with $E_{\rm i}=12$~meV on the CNCS instrument. Figure~\ref{CNCS_slices}(d) shows the energy cuts taken along the $(101)$ direction at $T=2$~K and $T=15$~K.
In order to extract the gap value we took the inflection points, as shown in Fig.~\ref{CNCS_slices}(d), and found $\Delta=4.03(5)$~meV for 2~K and $\Delta=4.88(5)$~meV for 15~K.

\subsection{Low-energy INS data}

\begin{figure}[tb]
    \includegraphics[width=1\columnwidth]{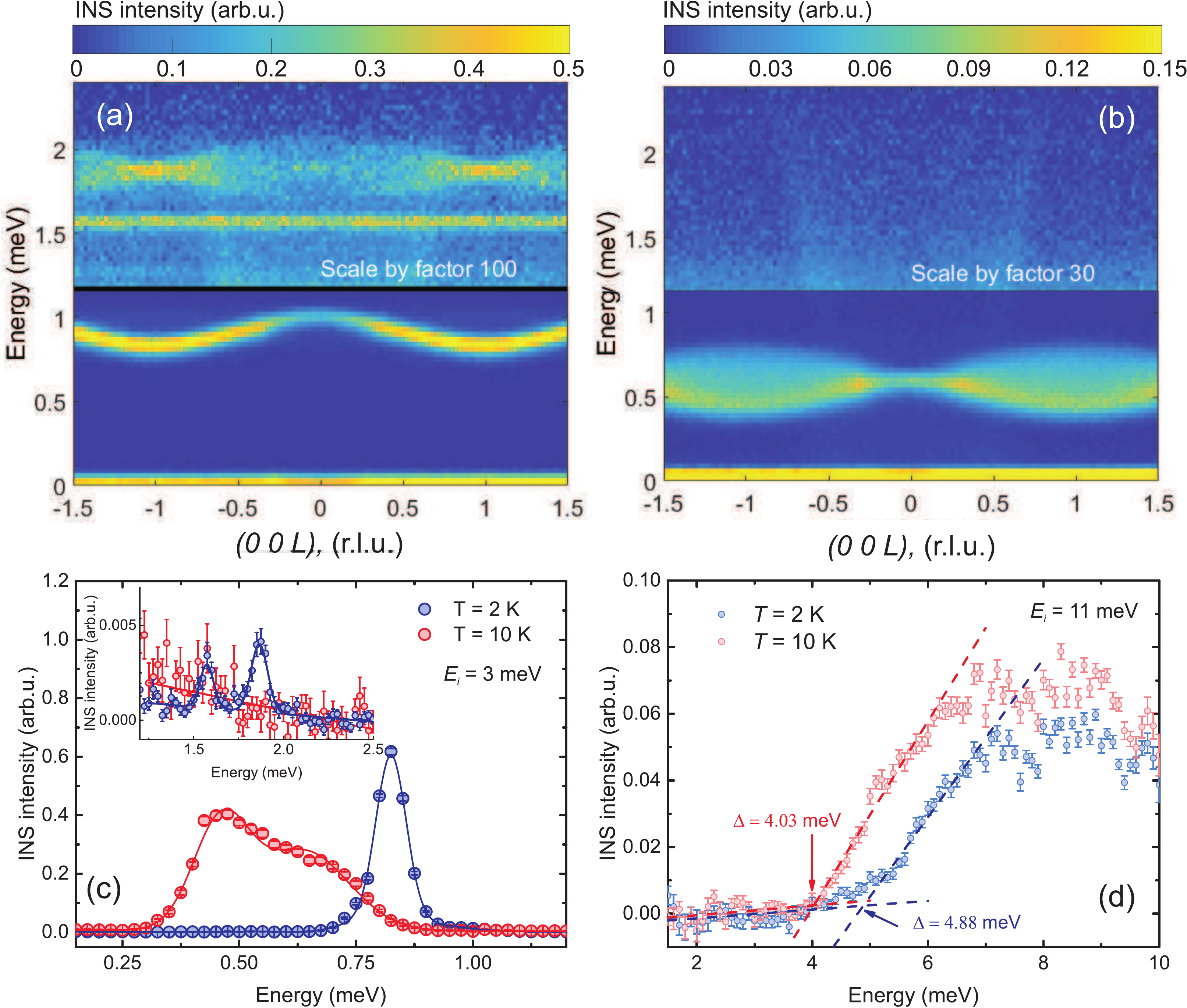}
    \caption{~Low-energy excitation spectra of YbFeO$_3$ at $T=2$~K (a) and $T=10$~K (b) taken at CNCS. Energy slices are taken along the $(00L)$ direction with $(0K0)$ and $(H00)$ integrated over the range [-0.5, 0.5]. The intensity of the upper part of the panels has been scaled $\times 100$ and $\times 30$ to make two-kink excitations visible. The shadow in the center is due to direct beam. (c) Energy cuts through the (001) direction taken for the both temperatures. Solid lines show results of fitting with one and two Gaussians for $T=2$~K and $T=10$~K, respectively. The intensity of the $T=10$~K spectrum was scaled with a factor of 3. Inset: zoom of the energy cut at [1.2, 2.5]~meV range, showing 2-kink excitations at 2~K. (d) Energy cuts at $\mathbf{Q}=$~(101), taken at $T=2$ and 10~K to show the gap in the excitation spectrum of Fe$^{3+}$ magnons.}
    \label{CNCS_slices}
\end{figure}

According to specific-heat measurements published previously~\cite{Moldover}, the Yb$^{3+}$ ground-state doublet has a splitting of 1~meV, therefore, in order to investigate the spin dynamics of the Yb subsystem we performed measurements on the CNCS instrument with $E_{\rm i}=3$~meV in the $(H0L)$ and $(0KL)$ scattering planes. Experimentally observed intensity maps, $I(\mathbf{Q},E)$ along the $(00L)$ direction are shown in Fig.~\ref{CNCS_slices}(a,b) for the temperatures below and above the SR transition. The excitation spectrum  at $T=2$~K is dominated by a high-intensity sharp mode, which disperses only along the ${Q}_{L}$ direction. At the zone center this mode peaks at $E_1\approx~1$~meV.
We also observe a weak dispersionless excitation at $E_2\approx 1.5$~meV and a continuum centered at $E_3~\sim~1.8$~meV with dispersive boundaries and a bandwidth of $\Delta{}E\approx0.3$~meV at the zone center.
Above $T_{\mathrm{SR}}$, a different spectrum emerges.
A bow-tie-shaped continuum arises at $E\approx0.6$~meV with a sharp mode observed at the lower boundary. The low-intensity excitation $E_2$ and the continuum $E_3$, present at $T=2$~K, totally disappear. Fig.~\ref{CNCS_slices}(c) shows energy cuts taken at $\mathbf{Q}=(001)$. One can see that all $E_1$, $E_2$ and $E_3$ peaks, observed at $T=2$~K, could be described with a single Gaussian function, whereas a cut, taken through the center of the continuum at $T~=~10$~K, consists of two peaks: relatively narrow, intense peak centered at $E~=~0.47$~meV and a second broad peak at $E~=~0.63$~meV.
All observed excitations have negligible dispersion along other directions (see additional Figures in~\cite{SI}), indicating that the Yb moments form weakly coupled spin chains running along the $c$-axis despite the three-dimensional perovskite structure, in a similar fashion as it was proposed for isostructural YbAlO$_3$~\cite{radhakrishna1981antiferromagnetic}.

Moreover, in both spectra taken above and below $T_{\mathrm{SR}}$ we observed a second ``shadow'' mode with similar dispersion, but shifted periodicity. It has no intensity at $Q_K~=~0$, but becomes visible at higher $Q_K$.
We describe the spectrum taken at 2~K using a LSWT calculation and show, that this mode is associated with the buckling of the Yb chains along the $b$ axis (details are presented in~\cite{SI}).

\begin{figure}
\includegraphics[width=1\columnwidth]{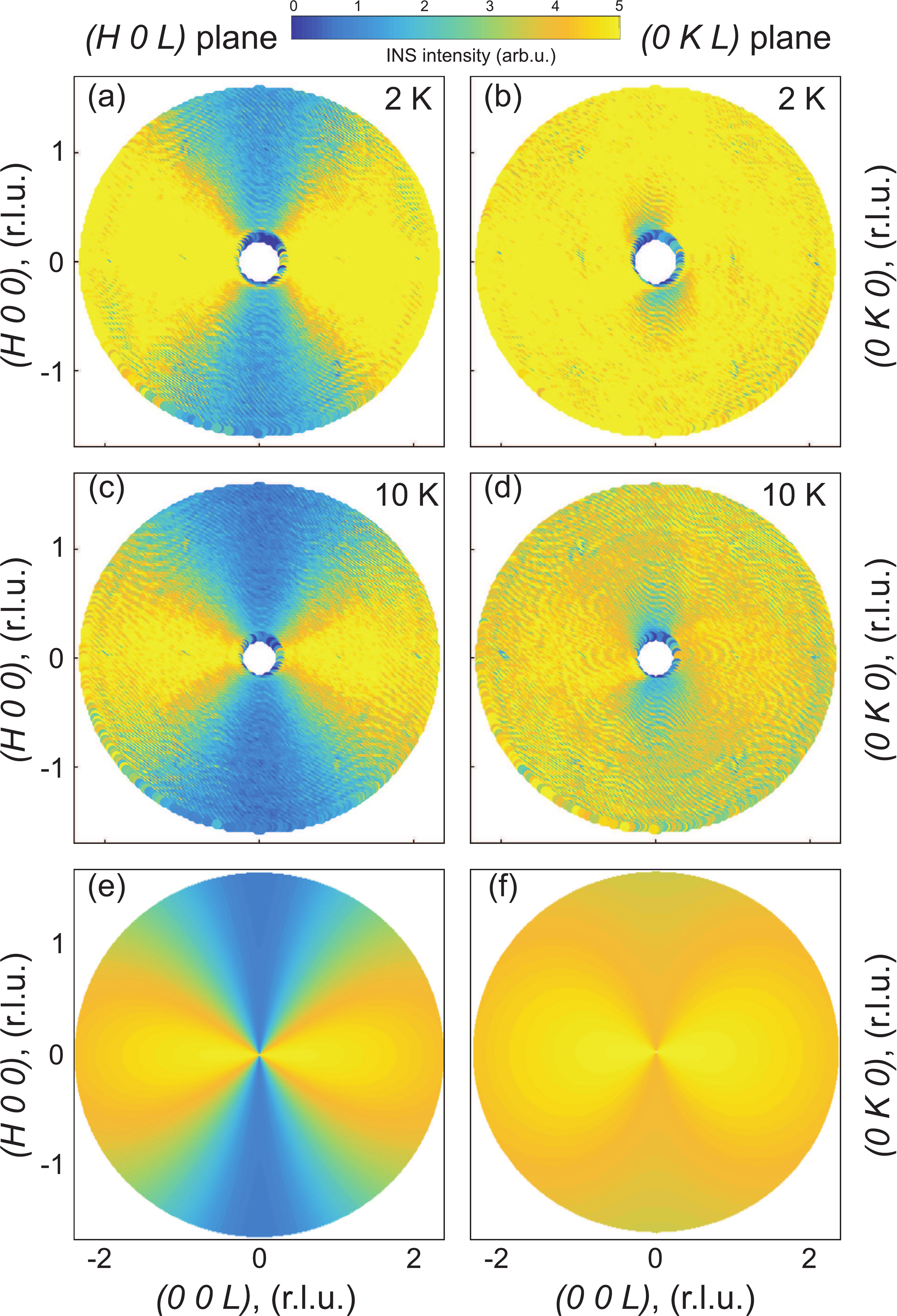}
    \caption{~Measured (a-d) and calculated (e,f) constant energy plots in the $(H0L)$ (left column) and $(0KL)$ (right column) scattering plane at $T=2$~K (a,b) and at $T=10$~K (c,d). The scattering intensity was integrated within $E = [0.8, 1.2]$~meV and $[0.4, 0.8]$~meV for $T=2$~K and 10~K, respectively. }
  \label{CNCS_const_E}
\end{figure}

\subsection{Effect of the polarization factor on the INS spectra}
\label{polarization}
Before one can start a discussion or some quantitative analysis of the spin dynamics in magnetic materials, it is very important to establish the static magnetic structure, which, in the general case, could be obtained from neutron diffraction measurements.
Magnetic structure of the Fe subsystem was determined and published for both $\rm \Gamma4$ and $\rm \Gamma2$ magnetic configurations~\cite{plakhty1983neutron, koehler1960neutron, White, bozorth1958magnetization}.
To the best of our knowledge there is no magnetic ordering of the Yb sublattice down to $T~\approx~100$~mK.
Therefore we can only discuss the preferred orientation of Yb moments, which can be caused by both Yb single-ion anisotropy due to the CEF~\cite{Wu2017DyScO3} and Yb-Fe interactions, including both dipole-dipole and exchange terms.
Previous measurements of YbFeO$_3$ using M\"{o}ssbauer spectroscopy~\cite{Davidson}, as well as theoretical work by Yamaguchi~\cite{Yamaguchi}, concluded that the Yb moments are strongly coupled to the Fe subsystem and therefore, Yb spins rotate from the $a$ to $c$ axis at $T_{\mathrm{SR}}$.
Our qualitative analysis of the polarization of INS presented below disagrees with this conclusions.

The polarization factor of neutron scattering affects the final scattering intensity, because only magnetic moment components perpendicular to the scattering vector $\mathbf{Q}$ contribute to the magnetic cross section.
The longitudinal component $S^{zz}$ that is mostly contributed from the moments along the $Q_i$ ($i=H, K, L)$ direction should follow the polarization factor:
\begin{eqnarray}
    p_i = 1-\frac{(Q_i)^{2}}{(Q_{H})^{2}+(Q_{K})^{2}+(Q_{L})^{2}}.
    \label{Polarization1}
\end{eqnarray}
Taking into account the form factor of the magnetic ion $|f(\mathbf{Q})|$, the integrated scattered intensity has the $\mathbf{Q}$-dependence
\begin{eqnarray}
  \int I(\mathbf{Q}, E)  dE \propto |f(\mathbf{Q})|^{2}\cdot{}p_i \nonumber\\
  \label{Polarization2}
\end{eqnarray}
Eq.~(\ref{Polarization2}) predicts a cone-shaped scattering, and the strongest intensity is recorded $\mathbf{Q}\perp\mathbf{Q}_i$.

The CEF lifts the degeneracy of the $4f^{13}$ electronic configuration of Yb$^{3+}$ into four Kramers doublets. Since the spin dynamics at the energy scale of $E\approx1$~meV is associated with fluctuations of Yb moments, the low-energy INS should reflect the wavefunctions anisotropy of the Yb ground-state doublet.
The CEF is controlled by the near neighbor coordination, which is little affected by isostructural substitution of rare-earth ions in the $R$FeO$_3$ family. 
Therefore, in order to estimate an effect of crystal field and the ground state wavefunctions of Yb$^{3+}$, we used CEF parameters determined for NdFeO$_3$~\cite{Przenioslo}.
We found that the Yb moments have a strong Ising-like anisotropy and lie in the $ab$ plane forming an angle within $\alpha=\pm21^{\circ}$ to the $a$ axis (See~\cite{SI} for details).
Figure~\ref{CNCS_const_E} presents $\mathbf{Q}$-dependencies of INS scattering taken within $(H0L)$ and $(0KL)$ planes for temperatures above and below the SR transition.
At both temperatures, $T=2$~K and 10~K, the INS intensity integrated over the range of Yb-spin excitations has a strong anisotropy in the $(H0L)$ plane, whereas the signal in the $(0KL)$ plane is almost isotropic.
In order to describe such scattering intensity we calculated $\mathbf{Q}$-dependences of the INS intensity in the both $(H0L)$ and $(0KL)$ planes, assuming that the Yb moments lie in the $ab$ plane with $\alpha=\pm21^{\circ}$ degree to the $a$ axis.
In this case, Eq.~\ref{Polarization1} describing a polarization factor of the neutron scattering can be rewritten in a following forms:
\begin{eqnarray}
 p_{(H0L)} = \frac{(Q_L)^2+\cos^2\!\alpha(Q_H)^2}{(Q_{H})^{2}+(Q_{L})^{2}},\\
 p_{(0KL)} = \frac{(Q_L)^2+\sin^2\!\alpha(Q_K)^2}{(Q_{K})^{2}+(Q_{L})^{2}},
 \label{polarization3}
\end{eqnarray}
for $(H0L)$ and $(0KL)$ scattering planes, respectively.
Because we assumed, that the magnetic moments of the Yb lie close to the $[100]$ direction, INS intensity, calculated for the $(H0L)$ plane, has a strong anisotropy [Fig.~\ref{CNCS_const_E}(e)].
On the other hand, the polarization factor of the INS scattering in the $(0KL)$ plane has only a weak $\mathbf{Q}$ dependence with maximums of the intensity along the $(00L)$ direction as shown in Fig.~\ref{CNCS_const_E}(f).
At both temperatures, $T=2$ and 10~K, the INS intensity integrated over the range of Yb spin excitations is qualitatively consistent with the calculations, as one can see in Fig.~\ref{CNCS_const_E}(a-f).
Thus, at both temperatures below and above SR transition, fluctuations we observed are dominated by the longitudinal component $S^{zz}$ along the easy axis of Yb magnetization.

Note that the INS intensity of the first CEF excitation at $E\approx20$~meV is concentrated along the $(100)$ direction (see Fig.~\ref{SEQUOIA_const_E}(c,d)), perpendicular to the low-energy $E \approx 1$~meV excitation.
A strong similarity of Yb excitations at $T=2$ and 10~K confirms that the magnetic anisotropy and the symmetry of wavefunctions of Yb$^{3+}$ remains the same despite the SR transition, contrary to previous reports~\cite{Davidson,Yamaguchi}.
This fact is also in a good agreement with the magnetization data as well as the results of the CEF calculations for YbFeO$_3$~\cite{SI}, showing that the ground state doublet of Yb has a strong Ising-like anisotropy with easy-axis lying close to the $a$ axis, whereas the first excited doublet, which has a different symmetry, is located at the energy transfer of $\sim~20$~meV, and therefore, can not influence the low-temperature magnetic properties.

\section{Zero field measurements: Interpretation}
\subsection{Magnetic Hamiltonian of YbFeO$_3$}

Coming to the quantitative description of the experimental results, we want to point out that in the general case Hamiltonian describing the spin dynamics of YbFeO$_3$ for both rare-earth and Fe sublattices should take into account three different terms:
\begin{eqnarray}
 \mathcal{H} = \mathcal{H}_\text{Fe-Fe} + \mathcal{H}_\text{Yb-Yb} + \mathcal{H}_\text{Fe-Yb},
 \label{Hamiltonian_Fe+Yb}
\end{eqnarray}
where the first two terms describe exchange interactions and single-ion anisotropies within Fe and Yb subsystems, respectively.
The third term is an effective interaction between the Fe and Yb subsystems, including both dipole-dipole and exchange terms.
A few decades ago Yamaguchi proposed and analyzed a model, which took into account all symmetric and antisymmetric exchange interactions within the Fe sublattice as well as interactions between Fe and $R$ sublattices, whereas the interactions and anisotropy within the $R$ sublattice were neglected ~\cite{Yamaguchi}.
The excitation spectrum of this model consists of a number of entangled collective Fe-$R$ spin-wave modes, as was shown for many other compounds with magnetic interaction between different sublattices~\cite{nakajima2011magnons, hayashida2015magnetic, golosovsky2017spin, TymoshenkoOnykiienko17, PyttlikNd2CuO4, princep2017Garnet}.

In contrast, for both temperatures, below and above the SR transition, in our experimental spectra we were able to separate two groups of collective excitations with rather different energy scales: (i) quasi-1D mode, caused by Yb-Yb exchange along the $c$ axis at $E\approx1$~meV [see Fig.~\ref{CNCS_slices}(a,b)] and (ii) gapped high energy spin-waves modes [see Fig.~\ref{SEQUOIA_slices}(a,b)], similar to other orthorhombic orthoferrites and associated with Fe-Fe exchange and Fe single-ion anisotropy~\cite{Hahn, park2017low}.
Therefore, in order to phenomenologically describe main features of the observed spin dynamics we decouple the Yb and Fe subsystems and construct the effective Heisenberg-like spin Hamiltonians for each of them separately.

Previously it was shown, that in the $R$FeO$_3$, influence of the Fe subsystem on the $R$ moment can be described in terms of an effective field~\cite{BelovBook, belov1974new, belov1979magnetic}, and here we followed the approach of the ``modified mean-field theory'', recently developed for $R$FeO$_3$~\cite{bazaliy2004spin, bazaliy2005measurements, tsymbal2007magnetic}.
Bazaliy \textit{et al.} analyzed a free energy functional of ErFeO$_3$~\cite{bazaliy2004spin}.
They assumed that the ordered Fe subsystem polarizes nearly paramagnetic, strongly anisotropic moments of $R$-ions by an internal molecular field $\mathbf{H}^{\rm Fe}$.
In this model, one can take into account the influence of ordered Fe moments on the Yb subsystem with a simple Zeeman term and write down the magnetic Hamiltonian for the Yb moments in a form:
\begin{eqnarray}
 \mathcal{H}_{\rm Yb} =  \sum_{l,m,i} B^l_mO^l_m(\mathbf{J}_i^{\rm Yb}) + J \sum_{\langle{}i,j\rangle} \ {\mathbf{J}^{\rm Yb}_{i} \cdot \mathbf{J}^{\rm Yb}_{j}}  + {\mathbf{H}}^{\rm Fe}\sum_i \mathbf{J}_i^{\rm Yb},
 \label{YbHam1}
\end{eqnarray}
where the first term is an one-site CEF Hamiltonian in Stevens notations~\cite{stevens1952matrix, hutchings1964point}, the second term is the Yb-Yb intersite Heisenberg exchange interaction, and the third term represents an influence of the Fe molecular field on the Yb magnetic subsystem.

Now, let us focus on the choice of the model Hamiltonian for description of the magnetic structure and spin dynamics of the Fe subsystem.
Without taking into account the Yb subsystem, it could be written in the following form:
\begin{eqnarray}
 \mathcal{H}_{\rm Fe} =  \sum_{\langle{}i,j\rangle} \ {\mathbf{S}^{\rm Fe}_{i} \cdot J_{ij} \cdot \mathbf{S}^{\rm Fe}_{j}}  - \sum_i \mathbf{S}^{\rm Fe}_i \cdot K_i \cdot \mathbf{S}^{\rm Fe}_i,
 \label{FeHam1}
\end{eqnarray}
where $J_{ij}$ is a $3\times3$  matrix, containing both symmetric and Dzyaloshinskii-Moriya (DM) Fe-Fe intersite exchange interactions and $K_i$ is a diagonal $3\times3$ matrix describing the effective single-ion anisotropy of the Fe moments.
Due to the orthorhombic symmetry of the Fe environment, the anisotropy matrix $K_i$ contains two nonequivalent constants $K_a$ and $K_c$.
In this Hamiltonian, the first term dictates an overall shape and maximum energy of the Fe excitations, the anisotropy determines a magnetic ground state~\cite{Belov} and gives rise to the gap in the Fe magnon spectrum~\cite{Hahn}.
In $R$FeO$_3$ with non-magnetic $R$-ions, a dominating $K_a$ stabilizes the $\rm \Gamma4$ phase, whereas in compounds with magnetic $R$, the $R$-Fe interaction induces renormalization of the effective anisotropy constants.
At $T \approx T_{\mathrm{SR}}$, $K_a$ and $K_c$ become approximately equal, and the term $\propto{}(S^z)^4$ controls the rotation of the Fe spins~\cite{Belov}.
Below the SR transition $K_c>K_a$ stabilizes the $\rm \Gamma2$ phase.
Having in mind that (i) the high-energy magnons in YbFeO$_3$ do not change through the SR transition and (ii) there are no collective Fe-Yb modes, we describe the evolution of the magnetic ground state and high-energy spin dynamics of the Fe subsystem, introducing a temperature dependency of the effective anisotropy constants $K_a^{\prime}(T)$ and $K_c^{\prime}(T)$ due to the Yb-Fe interaction.
Note, that the $K_c^{\prime}$ dominates in $\rm \Gamma2$, while $K_a^{\prime}$ dominates in the $\rm \Gamma4$ phase.
In the supplementary information~\cite{SI} we present a detailed analysis of the free energy functional of YbFeO$_3$ at temperatures close to $T_{\mathrm{SR}}$ and clarify, why the $R$-Fe exchange interaction leads to the SR transition and induces renormalization of the effective anisotropy constants.

We should point out that this is an entirely phenomenological approach, which, however, describes the details of the magnetic behavior of YbFeO$_3$ as well as most of the features of the observed spin dynamics.
Construction of the microscopically full magnetic Hamiltonian without decoupling of the Fe and Yb subsystems goes far beyond the scope of our work, but we hope that the results of our study will motivate further theoretical work on the unconventional spin dynamics in YbFeO$_3$ and explain the microscopic mechanism of the $R$-Fe interaction in rare-earth orthoferrites.

\subsection{Linear spin-wave model for the Fe magnons}

\begin{figure}
\includegraphics[width=0.8\columnwidth]{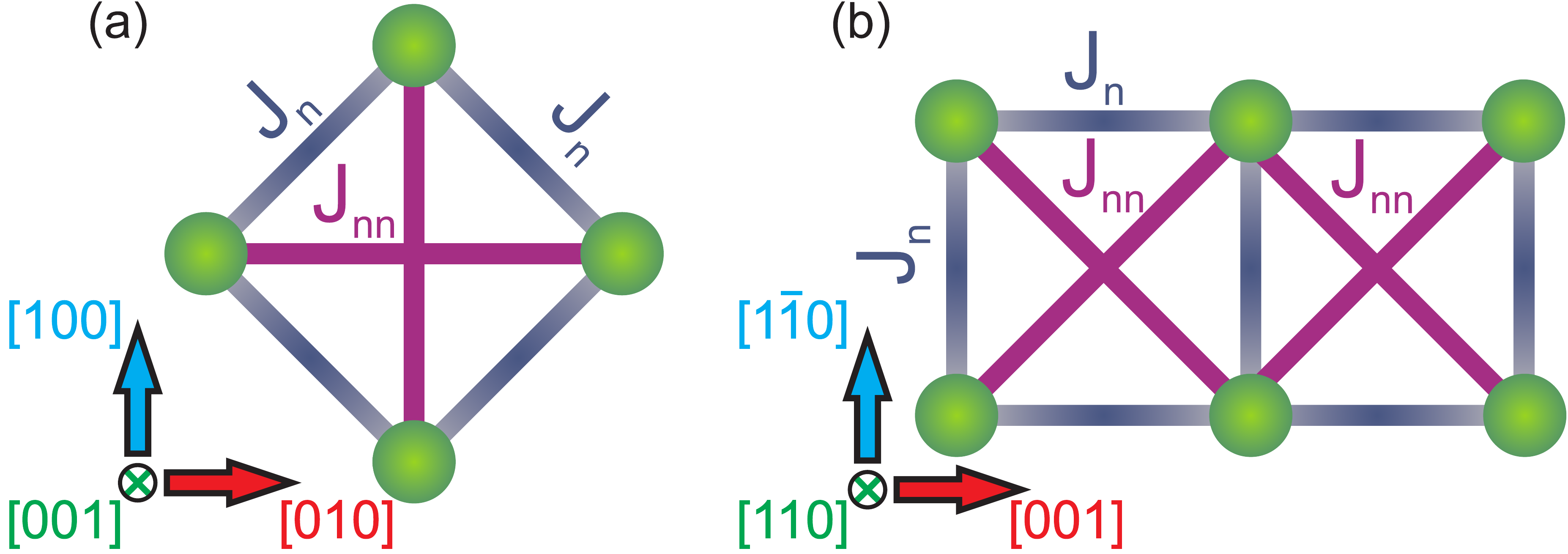}
    \caption{~Sketch of the Fe-Fe exchange paths in [001] (a) and [110] (b) planes.}
  \label{Exchange_path}
\end{figure}

As the first step, we focus on high-energy spin dynamics of the Fe subsystem.
Recently, a general Hamiltonian (Eq.~\ref{FeHam1}), describing the magnetic properties of the Fe subsystem, was written in a following form, in order to describe spin structure and dynamic properties of the isostructural YFeO$_3$~\cite{Hahn}:
\begin{eqnarray}
 \mathcal{H_{\rm Fe}} =J_{nn} \sum_{\langle i,j \rangle} \mathbf{S}_i\mathbf{S}_j + J_{nnn} \sum_{\langle i,j \rangle'} \mathbf{S}_i\mathbf{S}_j  \nonumber\\
 - D_1 \sum_{\substack{R_j=R_i\\+a(x \pm y)}} \mathbf{S}_i \times \mathbf{S}_j \nonumber \\
 - D_2 \sum_{\substack{R_j=R_i\\+a(x \pm y)}} \mathbf{S}_i \times \mathbf{S}_j \nonumber \\
 - K_a^{\prime}\sum_{i} (S_i^x)^2 - K_c^{\prime}\sum_{i} (S_i^z)^2.
 \label{LSWT}
\end{eqnarray}
It contains two isotropic exchange interactions between nearest-neighbor and next-nearest neighbor Fe ions (see Fig.~\ref{Exchange_path}), two DM exchange interactions within the $ab$-plane, and two effective easy-axis anisotropy constants $K_a^{\prime}$ and $K_c^{\prime}$.
As we discussed above, in order to take into account Yb-Fe exchange interaction and stabilize the correct ground state, either $\rm \Gamma4$ or $\rm \Gamma2$, we assume that the effective $K_a^{\prime}$ and $K_c^{\prime}$ are changing with temperature.
A large gap in the magnon spectra $\Delta~\approx~4$~meV [see Fig.~\ref{CNCS_slices}(d)], observed at both temperatures, $T>T_{\mathrm{SR}}$ and $T<T_{\mathrm{SR}}$, indicates an easy-axis character of the dominating anisotropy constant.

In rare-earth orthoferrites, DM exchange interactions give rise to the canted magnetic structure and an optical magnon branch at $E~\approx~65$~meV~\cite{Hahn}.
However, effective values of the DM parameters are rather small and therefore, corresponding branches have a vanishingly small spectral intensity, so we could not observe them in our INS data. On the other hand, knowing the canting angles $\theta~=~0.35^{\circ}$ and $\phi~=~0.18^{\circ}$.

We would like to note that first two symmetric Heisenberg exchange interactions define the energy scale and overall shape of the magnon branches.
The dominating anisotropy constant determines the ground state ($\rm \Gamma4$ or $\rm \Gamma2$) and gives rise to the gap in the excitation spectrum.
The presence of the DM exchange leads to a spin canting of the Fe spins~\cite{plakhty1983neutron,Hahn}.
The DM terms and the second anisotropy constant play a minor role in the spectrum and their spectroscopic determination requires additional careful measurements~\cite{park2017low}.
Therefore, to reproduce the magnon excitations of the Fe subsystem we used Hamiltonian~(\ref{LSWT}), with $J_{nn}$, $J_{nnn}$, $K_a^{\prime}$ (for $T>T_{\mathrm{SR}}$) and $K_c^{\prime}$ (for $T>T_{\mathrm{SR}}$) as free parameters, whereas $D_1$ and $D_2$ constants were calculated from the canting angles and fixed for both temperatures.
In order to derive parameters from the experimental spectra, we fit the experimental data at 28 different points of $\mathbf{Q}$-space along nonequivalent directions and extracted the energy and intensities of the magnon mode. Then, we fitted these points to our model Hamiltonian using \textsc{SpinW} software~\cite{Toth}.
The best sets of exchange parameters for both phases are shown in Table~\ref{J_table}.
Calculated dispersion curves, shown in Fig.~\ref{SEQUOIA_slices}(a,b) as the white dashed lines, are in good agreement with experimental data.
\begin{table}[bth!]
\caption {~Parameters of the magnetic Hamiltonian~(\ref{LSWT}) derived in this work. All values are given in meV. }\label{J_table}
 \begin{ruledtabular}
 \begin{tabular}{lllllll}
 Magnetic phase            &    $J_{nn}$   &      $J_{nnn}$  &  $D_1$   &    $D_2$    &     $K^{\prime}_a$    &   $K^{\prime}_c$   \\
 \hline
 $\rm \Gamma2$ ($T=2$~K)   &   4.675       &       0.158     &  0.086   &    0.027    &      0        &    0.023  \\
 $\rm \Gamma4$ ($T=15$~K)  &   4.675       &       0.158     &  0.086   &    0.027    &      0.033    &    0      \\
\end{tabular}
\end{ruledtabular}
\end{table}

\subsection{Quantum quasi-1D excitations in the Yb subsystem}

Having described the high-energy magnetic excitations of the Fe sublattice, we now discuss the low-energy magnetic excitations of the Yb$^{3+}$ moments observed in YbFeO$_3$.
The CEF term in Hamiltonian~(\ref{YbHam1}) gives a large splitting of the $J~=~\frac{7}{2}$ multiplet of Yb$^{3+}$.
The energy gap between the ground state and the first excited doublet is $\Delta~=~20$~meV.
Details of the influence of the CEF term on the $J~=~\frac{7}{2}$ multiplet of Yb$^{3+}$ are discussed in~\cite{SI}.
Therefore, for the description of the low-energy spin dynamics we can take into account the ground-state doublet alone and use the pseudo-spin $S=\frac{1}{2}$ approximation.
As we mentioned above, nearest-neighbour Yb moments are coupled along the $c$ axis by an exchange interaction.
In a simple approximation, the influence of the Fe subsystem on the Yb ions could be taken into account via the effective molecular field, which is created by the Fe sublattice as was discussed previously.
We transform Eq.~(\ref{YbHam1}) into the one-dimensional XXZ $S~=~\frac{1}{2}$ Hamiltonian:
\begin{eqnarray}
 \mathcal{H_{\rm Yb}} = && J_z \sum_i \ {S^z_i S^z_{i+1}} + J_{xy} \sum_i \ {(S^x_i S^x_{i+1} + S^y_i S_{i+1}^y)}\ \nonumber\\
 && + \sum_i \mathbf{H_{\rm ef}} \cdot \mathbf{S}_i,
 \label{YbHam2}
\end{eqnarray}
where the first two terms correspond to the anisotropic exchange interaction between the nearest-neighbor Yb along $c$ axis, and the last term is an effective Zeeman term -- sum of the external field and the molecular field of the Fe subsystem.

At temperatures $T<T_{\mathrm{SR}}$ the net moment of the Fe subsystem is directed along the $a$ axis, as shown in Fig.~\ref{SEQUOIA_slices}(c), creating a \emph{longitudinal} field for Yb$^{3+}$ spins.
In order to describe the low-$T$ spectrum, we performed calculations of the eigenstates of Eq.~(\ref{YbHam2}) using the zero temperature exact diagonalization of a finite chain ($L=20$) with ALPS software~\cite{ALPS1, ALPS2}.
A cosine-shape dispersion of the lowest excitation with a maximum at the zone center suggests that the exchange interaction is antiferromagnetic~\cite{SI} and that the effective field $\mathbf{H}_{\rm ef}$ is large in comparison to $J_z$ and $J_{xy}$.
In this case all spins are parallel, $\langle S_n^z \rangle = S$~\cite{Mourigal}.
The excitation spectrum of such a fully polarized state is similar to that of a FM chain and was discussed in detail a few decades ago~\cite{Torrance595, Torrance587, Fogedby, Schneider}.
A single sharp mode with energy $E\approx1$~meV occurs due to scattering by a single-flip quasiparticle.
Besides, modes of an anisotropic FM or field-polarized AFM chain contain a two-kink bound state and a continuum consisting of pairs of independently propagating kinks.
We found the cross section of two-kink states to be about two orders of magnitude weaker than that for the single-flip excitation, in agreement with the theoretical prediction~\cite{Torrance595}.
In case of both $J_z>0$ and $J_{xy} > 0$, the calculated two-magnon bound state mode lies above the continuum, which contradicts our experimental data (Fig.~\ref{CNCS_slices}(a)).
Our data would be reproduced well for $J_z<0$ and $J_{xy}>0$.
However, the situation when a single exchange bond has both FM and AFM correlations between different spin components seems to be unrealistic.
Furthermore, the magnetic ground state of the isostructural YbAlO$_3$ was found to be AFM~\cite{radhakrishna1981antiferromagnetic}.
This question requires a separate theoretical study.

At temperatures $T>T_{\mathrm{SR}}$  the Fe net moment reorients along the $c$-axis, inducing a \emph{transverse} field for the Yb spins (see Fig.~\ref{SEQUOIA_slices}(e)).
However, at $T=10$~K the observed superposition of a bow-tie-shaped spinon-like continuum with a sharp excitation at the bottom (see Fig.~\ref{CNCS_slices}(b)), suggests that the Yb sublattice is in a partially polarized state, as if a weak longitudinal field were still present.
A weak coupling between the magnetic chains in the $ab$ plane, evident from a weak dispersion along $H$ and $K$ directions (see Figs.~5 and 6 in \cite{SI}), could be a possible explanation of the observed spectrum.
Such coupling in a first approximation can be replaced by an effective longitudinal mean-field \cite{Carr,Coldea}.
The spin-excitation spectrum in a skew ($H_x$, $H_z$) field is indeed characterized by a combination of a continuum due to scattering by pairs of kinks, which interpolate between regions with magnetization `up' and `down' and a sharp mode created by single spin-flip quasiparticles.
The finite temperature model of an XXZ chain is required to describe the details of the experimental spectra in this case.

\begin{figure*}
\includegraphics[width=0.9\linewidth]{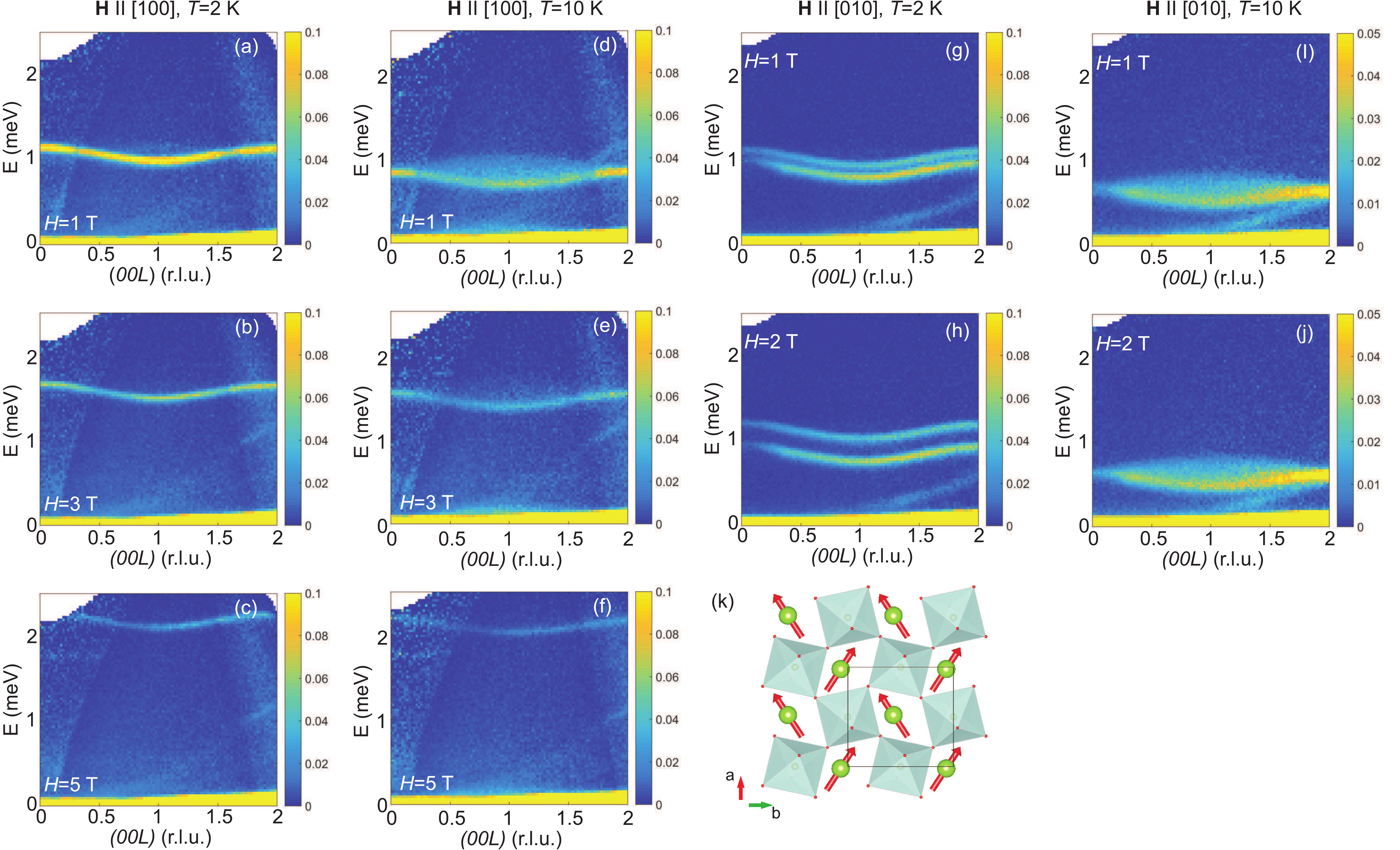}\vspace{3pt}
  \caption{
  ~Effect of magnetic field on the low-energy spin dynamics of YbFeO$_3$. Experimental spectrum along the $(00L)$ direction, measured at magnetic fields along the $a$ axis (a-c) and $b$ axis (g-j), at temperatures 2~K (a-c,g,h) and 10~K (d-f,i,j).  A visible linear diagonal line on the spectra is due to an instrumental effect.
  (k) Sketch of the field-induced magnetic structure of Yb moments below $T_{\mathrm{SR}}$.
  }
  \label{CNCS_magnetic}
\end{figure*}

\section{Magnetic field effect on the low-energy spin dynamics}

In previous sections we assumed that the influence of the ordered Fe subsystem on the Yb subsystem can be taken into account via the effective molecular field, which rotates from the $c$ to $a$-axis at $T_{\mathrm{SR}}$.
In this section we present the results of INS measurements with magnetic field applied along all $a$, $b$ and $c$ axes of the orthorhombic YbFeO$_3$ and show, that the effect of the external magnetic field on the spin dynamics is similar to that of the internal Fe-induced field. The results of the measurements for the $\mathbf{H}\parallel{}[100]$ and $\mathbf{H}\parallel{}[010]$ are summarized in Fig.~\ref{CNCS_magnetic}.

First of all, let us consider the low-temperature ($T<T_{\mathrm{SR}}$) spectra under the magnetic field along the $a$ axis, Fig.~\ref{CNCS_magnetic}(a-c).
In this case, Yb spins are already polarized along the easy $a$ axis even without an external magnetic field.
The external field leads to further Zeeman splitting of the ground state, whereas the total INS intensity of the excitation is decreasing.

At $T=10$~K, YbFeO$_3$ is in the $\rm \Gamma2$ phase, and the net moment is directed along the $c$ axis. Application of the magnetic field $\mathbf{H}\parallel{}[100]$ at this temperature has a dual effect: (i) it polarizes the Yb subsystem and (ii) induces a SR transition of Fe-moments $\rm \Gamma4 \rightarrow{} \Gamma2$. According to the specific-heat measurements~\cite{SI}, at $T=10$~K such a SR transition takes place at $H\approx4.3$~T.
In our INS data (see Fig.~\ref{CNCS_magnetic}(d-f)) we observe Zeeman splitting, whereas the continuum, dominating at zero field, is rapidly suppressed and becomes undetectable already at $H=3$~T.
At $H=5$~T, above the field-induced SR transition, the magnetic phase $\rm \Gamma2$ is stabilized.
The spectra at both temperatures, $T=2$ and 10~K, become identical.
Assuming the linear dependence of the energy splitting within the low-temperature $\rm \Gamma2$ phase, we calculated an effective $g$-factor $g^{\rm \Gamma2}_a=4.135$.

In contrast to the relatively simple case of $\mathbf{H}\parallel{}[100]$, a magnetic field applied along the $b$ axis qualitatively changes the excitation spectra. At temperatures below $T_{\mathrm{SR}}$ [Fig.~\ref{CNCS_magnetic}(g,h)] the single-particle mode splits into two parallel modes, whereas above $T_{\mathrm{SR}}$ magnetic field up to 2~T has a minor effect on the spectra [see Fig.~\ref{CNCS_magnetic}(i,j)].
Later, a special sample holder was constructed and used for the experiments with $\mathbf{H}\parallel{}[001]$.
According to the our model, below $T_{\mathrm{SR}}$ the Yb moments have an Ising-like anisotropy, lie in the $ab$ plane with $\alpha \approx \pm21^{\circ}$ to the $a$ axis and are fully polarized by the molecular field of the Fe subsystem. Schematically, molecular-field-induced magnetic structure of the Yb subsystem below $T_{\mathrm{SR}}$ is shown in Fig.~\ref{CNCS_magnetic}(k).
Application of a magnetic field along the $b$ axis lifts the degeneracy between neighbor magnetic chains, increasing the energy of fluctuations with the positive Yb moment projection on the $b$ axis, $\alpha=+21^{\circ}$, and decreasing the energy for the opposite direction, $\alpha=-21^{\circ}$. A further increase in field would suppress the energy of the lower mode down to zero with a simultaneous polarization of Yb moments along the $b$ axis.

In YbFeO$_3$, Yb moments are coupled in chains running along the $c$ axis, creating the dispersion along the $(00L)$ direction.
To apply a magnetic field along the $c$ axis, an experimental arrangement with {\em horizontal} field is preferred, since only magnetic moment components perpendicular to the scattering vector $\mathbf{Q}$ contribute to the magnetic cross section, as we discussed in Section~\ref{polarization}.
Therefore, for INS measurements in this geometry we oriented the sample in the $(H0L)$ scattering plane and used the triple-axis FLEXX instrument with the horizontal cryomagnet HM-1.
However, due to the instrument restrictions (dark angles of the magnet) we were limited with the $\mathbf{Q}$-range from (0~0~0.5) to (0~0~1.1) for $k_{\rm f}=1.3$~\AA$^{-1}$.

\begin{figure}
\includegraphics[width=\linewidth]{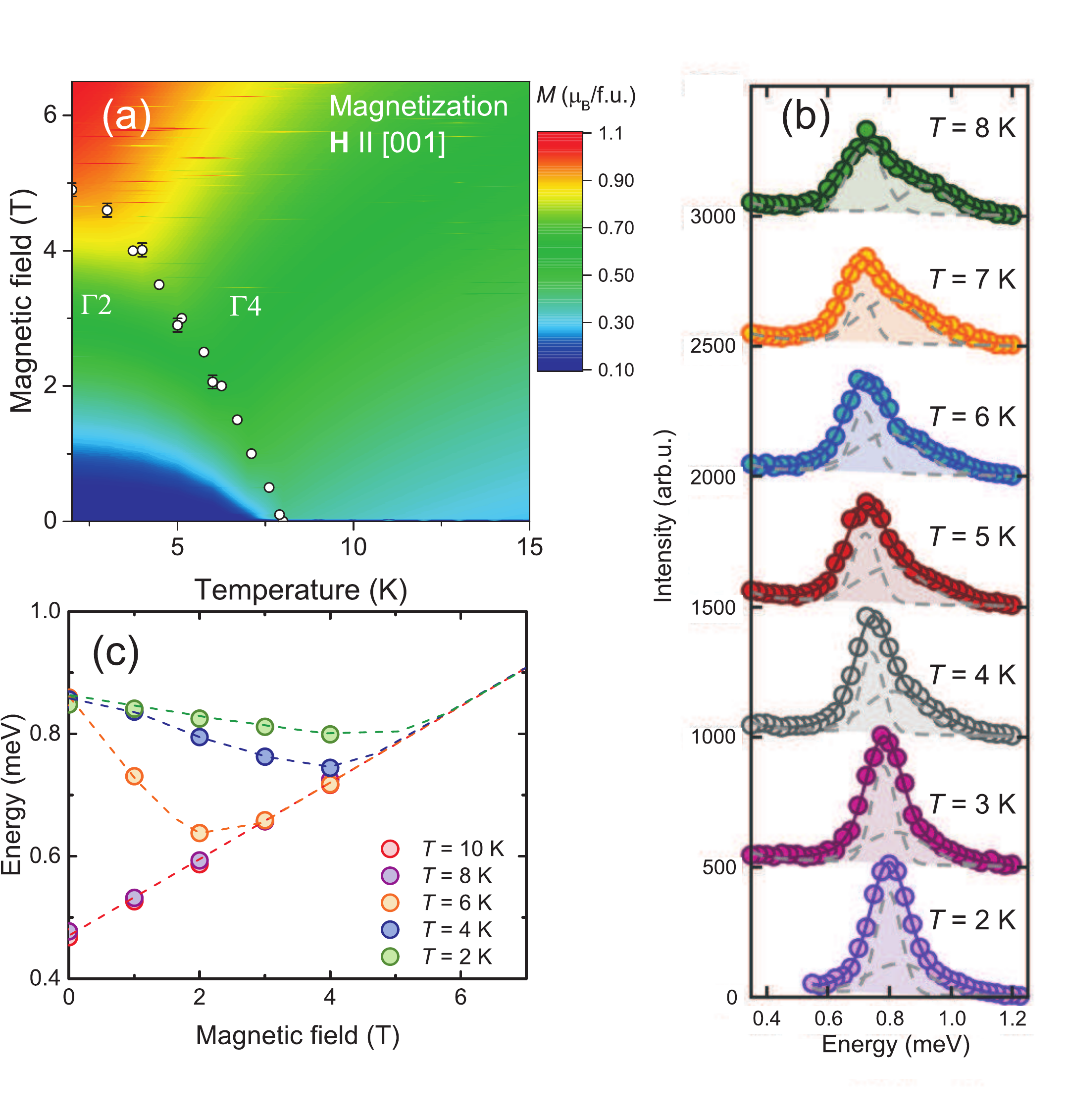}\vspace{3pt}
  \caption{~Effect of magnetic field along the $c$ axis on the low energy spin dynamics of YbFeO$_3$.
  (a) Magnetic-field--temperature phase diagram of YbFeO$_3$ taken at $\mathbf{H}$ applied along the $c$-axis.
  Color-plot shows magnetization data.
  (b) Energy scans, measured on the FLEXX instrument at $H=4$~T and $\mathbf{Q}=$~(001) at various temperatures. Solid line is an overall fit of the magnetic signal. Dotted lines represent two Gaussian functions, used for the fitting.
  (c) Magnetic field dependence of the ``main mode'' peak as a function of magnetic field. Dotted lines are drawn to guide the eyes.
}
  \label{Flexx}
\end{figure}

The magnetic field -- temperature phase diagram of YbFeO$_3$ reconstructed from the magnetic measurements is shown in Fig.~\ref{Flexx}(a).
One can see that the low-$T$ phase $\rm \Gamma2$, where the weak net moment of the Fe subsystem is aligned along the $a$ axis, could be suppressed by the magnetic field along the $c$-axis. The critical field $H_{\rm crit}^{\rm \Gamma2\rightarrow\Gamma4}$ gradually increases with the temperature decreasing.

Inelastic spectra taken at $\mathbf{Q}=$~(001) and $H=4$~T are described by the combination of two modes, a resolution-limited intense peak (``main'' mode) and an additional broad peak at higher energy, see Fig.~\ref{Flexx}(b).
We use two Gaussian functions for fitting the spectral line-shape.
The low-temperature scans ($T=2$ and 3~K) in the $\rm \Gamma2$ phase show the largest contribution of the ``main'' mode.
The center of the second peak is located very close to the first one.
At $T=4$~K, a field-induced SR transition occurs.
The second peak shifts to higher energies and its intensity grows, whereas further increase in temperature has no major effect on the spectra.

Figure~\ref{Flexx}(c) shows magnetic field dependence of the ``main'' mode taken at different temperatures.
The spectra taken at $T=2$ and 4~K show that the excitation energy is always growing up in the $\rm \Gamma2$ phase.
However, we found different behavior of the ``main'' magnetic peak at $T=6,8$ and 10~K.
First, the excitation energy goes down until the critical field $H_c^{\rm \Gamma2\rightarrow\Gamma4}$ [see Fig.~\ref{Flexx}(a)], and it starts growing at higher fields.
Thus, in the $\rm \Gamma2$ phase, increasing field reduces the energy of the excitation, whereas in the $\rm \Gamma4$ phase excitation energy rises with the field.
We also calculated the effective $g$-factor for the $\rm \Gamma4$ phase, which was found to be $g^{\rm \Gamma4}_c=1.09$, almost 4 times smaller compared to a $g^{\rm \Gamma2}_a=4.135$.

\section{Discussion and conclusions}

A large number of independent parameters of the full microscopic spin Hamiltonian of YbFeO$_3$~\cite{Yamaguchi} makes the analysis ambiguous and complicated.
However, quantitatively, one can consider three energy scales $J_\text{Fe-Fe}\gg{}J_\text{Fe-Yb}>J_\text{Yb-Yb}$. Strong $J_\text{Fe-Fe}$ interaction induces an AFM ordering in the Fe subsystem with $T_{\mathrm{N}} \approx 600$~K and its manifestations are clearly seen in high-temperature magnetic susceptibility or specific-heat measurements, magnetic neutron diffraction and INS spectra. The intermediate strength Yb-Fe interaction induces a spontaneous SR transition $\rm \Gamma4\rightarrow\Gamma2$ at decreasing temperature, and can be extracted from the low-temperature magnetization and specific-heat measurements, but the presence of $J_\text{Yb-Fe}$ exchange does not introduce new collective Yb-Fe modes or hybridization. Finally, the weakest 1D Yb-Yb correlations create unusual low-energy excitation spectra, which include a two magnon bound state, ``shadow'' mode, a spinon continuum etc. Note that on one hand, details of the Yb-Yb correlations are hidden for the most of the experimental macroscopic probes by dominating $J_\text{Fe-Fe}$ and $J_\text{Fe-Yb}$ interactions. On other hand, an \textit{ab-initio} DFT calculation, which can be used to identify the 1D character of Yb correlations does also fail to capture weak Yb-Yb correlations, due to the low one-site symmetry of both magnetic ions and presence of a second magnetic subsystem with much larger exchange energy. Therefore, high-resolution cold-neutron spectroscopy is a unique probe, which can explore details of the spin dynamics in the Yb subsystem and it is not surprising, that despite more than 60 years of investigations of rare-earth orthoferrites~\cite{bozorth1958magnetization}, quasi-one-dimensional Yb-Yb correlations have never been observed.

The main aim of this work is to present an experimental observation of the decoupled spin dynamics of the Fe and Yb subsystems, coexisting on different energy scales and to give a phenomenological description of the observed spectra.
We constructed spin Hamiltonians for each magnetic subsystem separately. The key simplification was to treat Yb-Fe interaction in terms of an effective ``mean-field'' approximation, instead of constructing a combined microscopic Hamiltonian, which should include both magnetic subsystems, and, therefore, terms $\propto{}S^{\rm Fe}\cdot{}S^{\rm Yb}$

We show that the magnetic structure and spin-dynamics of the Fe subsystem can be well described using the semi-classical LSWT.
This model takes into account the nearest neighbor exchange interaction and assumes the dominating effective easy-axis anisotropy constants $K^{\prime}_a$ or $K^{\prime}_c$ for the $\rm \Gamma4$ or $\rm \Gamma2$ phases, respectively.

Because the low-energy excitations were found to have a dispersion along the $c$ axis only, we concluded that the Yb nearest neighbor AFM exchange interaction along the $c$ axis dominates the exchange interactions within the $ab$ plane, despite the 3D crystal structure of YbFeO$_3$.
For the description of the Yb dynamics we propose a 1D-XXZ $S=\frac{1}{2}$ (Eq.~\ref{YbHam2}) Hamiltonian with the additional Zeeman term describing the effective interaction with the Fe subsystem.
The calculated excitation spectrum is in a reasonable agreement with the low-temperature experimental spectrum, when the molecular field of the Fe subsystem is longitudinal to the easy-axis of the Yb moments (at $T<T_{\mathrm{SR}}$).
The observed spectrum consists of the sharp intense single-magnon mode and two multi-magnon excitations: the dispersionless two-magnon bound state and the two-magnon continuum.
At $T>T_{\mathrm{SR}}$ in the $\rm \Gamma4$ phase, the molecular field of the Fe subsystem is aligned along the $c$ axis and transverse to the easy-axis of Yb moments, which lies in the $ab$ plane, with $\alpha~=~21^{\circ}$ to the $a$ axis~\cite{SI}.
We found that the single particle mode is shifted down in energy and accompanied by a broad spinon continuum, as it was reported for many other $S=\frac{1}{2}$ 1D magnets~\cite{Mourigal, tennant1993unbound, zaliznyak2004spinons, Wu2016Orbital}.

We performed calculations of the eigenstate spectrum for the 1D XXZ model including the transverse field~(\ref{YbHam2}) but could not find any set of parameters, which satisfactorily describes the experimentally observed excitations~\cite{SI}.
The apparent reason for such disagreement is that the model Hamiltonian~(\ref{YbHam2}) is oversimplified and not sufficient to describe the details of the low-energy spin dynamics in YbFeO$_3$ at finite temperatures.
We assume three main approximations: i) We took into account Yb-Yb exchange interaction along the $c$ axis only; ii) the  $J=7/2$ multiplet of Yb$^{3+}$ was substituted by the two-level pseudo-$S=\frac{1}{2}$ system; iii) We considered Yb-Fe exchange interaction as an effective internal field, following~\cite{bazaliy2004spin, bazaliy2005measurements, tsymbal2007magnetic}. The two first approximations are based on the number of experimental facts: 1D dispersion of Yb excitations; broad maximum on the temperature dependent magnetic susceptibility of the YbAlO$_3$, associated with the 1D spin correlations~\cite{radhakrishna1981antiferromagnetic}; spinon-like excitations above $T_{\mathrm{SR}}$; the large CEF gap in the INS spectrum $\Delta=20$~meV.
The third approximation is a common simplification, used for systems with several magnetic sublattices, where one energy scale significantly exceed others~\cite{thalmeier1996Nd2CuO4, henggelerThalmeier1996Nd2CuO4, fabreges2008magneticYbMnO3}.

Besides, instead of temperature dependent dynamical spin susceptibility $\chi^{\prime\prime}(\mathbf{Q},\hbar\omega)$ measured at the INS experiment, we calculated zero temperature eigenstates of the spin Hamiltonian.
In the low-temperature case $T<T_{\mathrm{SR}}$, we have an energy hierarchy of $H^{\rm Fe}\gg{}J^\text{Yb-Yb}\gg{}T$, and the calculated spectrum is split into the series of well define modes as clearly seen in Fig.~7 in \cite{SI}.
Above the $T_{\mathrm{SR}}$, $H^{\rm Fe} \sim T>J$, and zero-$T$ calculations become inapplicable.
Finite temperature effects should be taken into account in order to describe the dynamical spin susceptibility.

In summary, we present a comprehensive INS study of the spin dynamics in YbFeO$_3$ at temperatures close to the SR transition and in magnetic fields applied along three crystallographic directions. We constructed an effective model describing spin dynamics and static magnetic structure of Fe moments for both temperatures above and below $T_{\mathrm{SR}}$ assuming the temperature dependence of the effective single-ion anisotropy constants $K^{\prime}_a$ and $K^{\prime}_c$. In the low-energy magnetic spectra we observed an unusual transition between two regimes of the quasi-1D Yb fluctuations, induced by the rotation of the Fe molecular field, which serves as an intrinsic ``tuning parameter''. Our model Hamiltonian describes the main features of the low-temperature spectrum, whereas for the correct description of the spectrum at $T>T_{\mathrm{SR}}$ further theoretical work will have to be done.
We leave several open questions here: 1) What is the origin of the quasi-1D behavior within the Yb subsystem? 2) How to describe the unusual Yb excitation spectrum at $T>T_{\mathrm{SR}}$ with coexisting spinon and magnon modes? 3) What is the correct microscopical approach to describe the Fe-Yb exchange interaction instead of the mean-field approximation? We hope that the presented INS data and intriguing underlying physical phenomena would motivate further theoretical studies on YbFeO$_3$ and renew the interest to the rich physics of rare-earth orthoferrites in general, along with other materials with a coexistence of several magnetic subsystems on different energy scales.

\section{Acknowledgments}
We would like to thank A.~Sukhanov, O.~Stockert and P.~Thalmeier for useful discussions.
This research used resources at the Spallation Neutron Source, a DOE Office of Science User Facility operated by Oak Ridge National Laboratory. Part of this work was supported by the U.S. Department of Energy, Office of Science, Basic Energy Sciences, Materials Sciences and Engineering Division. D.S.I. acknowledges funding by the German Research Foundation (DFG) through the Collaborative Research Center SFB~1143 at the TU~Dresden (project C03). S.E.N. acknowledges support from the International Max Planck Research School for Chemistry and Physics of Quantum Materials (IMPRS-CPQM). L.S.W. was supported by the Laboratory Directed Research and Development Program of Oak Ridge National Laboratory, managed by UT-Battelle, LLC, for the U.S. DOE. S.B. and S.A.G. are supported by BFFR, grant No F18KI-022

%

\onecolumngrid
\clearpage

\begin{center}
\textbf{SUPPLEMENTARY INFORMATION:\\
Decoupled spin dynamics in the rare-earth orthoferrite YbFeO$_3$:\\
Evolution of magnetic excitations through the spin-reorientation transition.}
\end{center}

\twocolumngrid
\setcounter{equation}{0}
\setcounter{figure}{0}
\setcounter{table}{0}
\makeatletter

\renewcommand{\figurename}{Fig. S}
\renewcommand{\bibnumfmt}[1]{[S#1]}
\renewcommand{\citenumfont}[1]{S#1}

\subsection{Sample preparation}

Polycrystalline YbFeO$_3$ was prepared by a solid state reaction.
The starting materials of Yb$_2$O$_3$ and Fe$_2$O$_3$ with 99.99\% purity were mixed and ground followed by a heat treatment in air at $1000 - 1250$~$^{\circ}$C for at least 70 hours with several intermediate grindings.
The phase purity of the resulting compound was checked with a conventional x-ray diffractometer.
The resulting powder was hydrostatically pressed into rods (8~mm in diameter and 60~mm in length) and subsequently sintered at 1400~$^{\circ}$C for 20 hours.
The crystal growth was carried out using an optical floating zone furnace (FZ-T-10000-H-IV-VP-PC, Crystal System Corp., Japan) with four 500~W halogen lamps as heat sources.
The growing conditions were as follows: the growth rate was 5~mm/h, the feeding and seeding rods were rotated at about 15~rpm in opposite directions to ensure the liquid's homogeneity, and an oxygen and argon mixture at 1.5~bar pressure was applied during growth.
The lattice constants in the $Pbnm$ space group at room temperature were $a = 5.282(1)$~\AA, $b = 5.596(1)$~\AA, and $c = 7.605(1)$~\AA.

Growth on seeds of YbFeO$_3$ single crystals from B$_2$O$_3$ - BaF$_2$ - BaO solvent was carried out in a vertical thermo-shaft furnace provided with resistive heating elements.
The furnace design is similar to that described by Bezmaternykh $et al.$ in~\cite{bezmaternykh1984growth}.
The flux melt was prepared by successive melting of oxide components.
The melt was heated 50--70$^\circ$C above the expected saturation temperature  and, after intensive stirring with a platinum mixer, allowed to soak for 8--12~h.
The criterion for homogeneous state of the flux melt was stability of the saturation temperature.
Having determined the saturation and spontaneous crystallization temperature, individual crystal seeds were allowed to nucleate and grow up to a size of approximately 1$\times$1$\times$1~mm$^3$ at a temperature which was 1-2$^\circ$C lower than of spontaneous crystallization.
After thorough visual examination of the facet quality, the crystal seeds mounted on platinum shields were fixed to the crystal holder.
The following flux-melt over saturation regime was determined from the consideration of the number of crystal seeds, melt mass, concentration of solute components, width of the metastable zone and optimized rate of the YbFeO$_3$ single crystal growth.
Within the accuracy of the x-ray fluorescent analysis (0.02 wt \%), barium has not been detected in the single crystals.

\subsection{Crystal Electric Field effect}
\label{sec_CEF}

\begin{figure}[tbh!]
    \includegraphics[width=1\columnwidth]{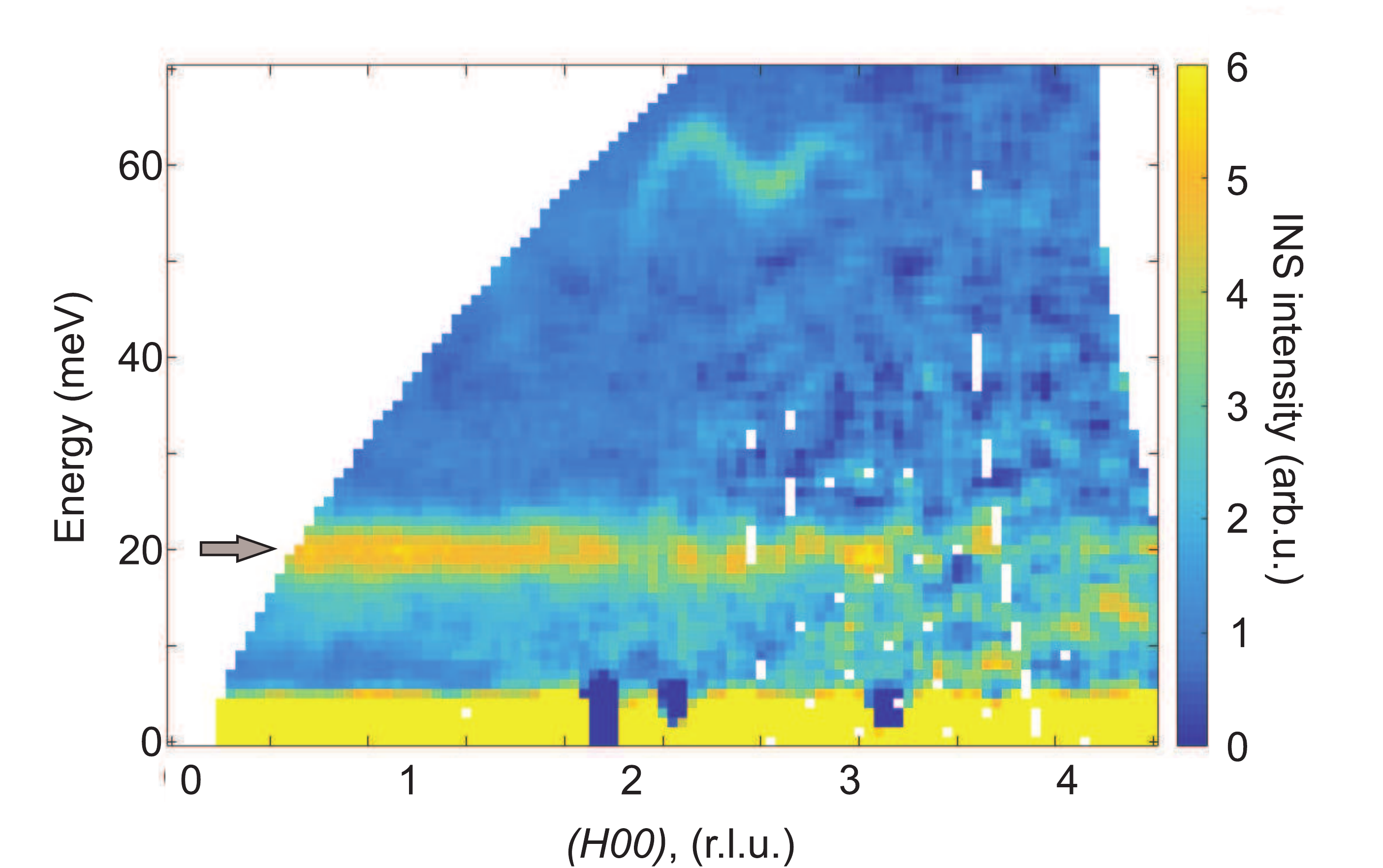}
    \caption{~INS spectrum taken at the SEQUOIA spectrometer at $T=5$~K, displaying CEF excitation in YbFeO$_3$ at 20~meV (shown bytra the arrow) as a function of energy and momentum through $(H00)$.}
    \label{fig:ins_cef}
\end{figure}
In the rare-earth lanthanides, the strength of the spin-orbit coupling significantly exceed the CEF effect, making the total angular moment $J$ a good quantum number. In the general case, for the Kramers ions with odd number of $4f$ electrons, the CEF Hamiltonian splits the $J$ multiplet into $\frac{2J+1}{2}$ doublets, with typical energy splitting of $\Delta\approx10$ -- 100~meV. Therefore, the low-temperature properties ($k_{\rm B}T\ll\Delta$) are dictated by low-lying doublet only and could be described with a pseudo-spin $S=\frac{1}{2}$ approximation, whereas real values of magnetic moments are absorbed into the effective anisotropic $g$-factor.

One experimental approach to explore CEF effects is to conduct measurements of the INS or optical spectra and to perform a subsequent fitting of the $B_l^m$ parameters to the energies and intensities of the observed transitions.
Our INS spectra exhibit only one dispersionless excitation at energy transfer $\Delta E = 20.0$~meV that can readily be identified as an Yb$^{3+}$ CEF transition from the ground state by the way in which its intensity varies with scattering vector (see Fig.~\ref{fig:ins_cef} and Fig.~1(a,b) in the main text). The CEF Hamiltonian for the $4f$ ion contains 15 independent coefficients $B_l^m$ for the orthorhombic symmetry, and therefore, an unambiguous interpretation of the CEF splitting scheme of Yb$^{3+}$ in YbFeO$_3$ is hardly possible.
Since the CEF is not controlled by the rare-earth ion $R$ itself but by the nearest neighbor coordination,
one can extrapolate the CEF level scheme of Yb$^{3+}$ using parameters known for another $R$.

A set of crystal field parameters $B_l^m$ was determined with good enough accuracy for NdFeO$_3$~\cite{Przenioslosi}.
For the extrapolation we use the standard equation
\begin{equation}
\frac{B_l^m({\rm Yb})}{\langle r^l({\rm Yb}) \rangle \theta_n({\rm Yb})} = \frac{B_l^m({\rm Nd})}{\langle r^l({\rm Nd}) \rangle \theta_n({\rm Nd})}
\label{eq:Blm}
\end{equation}
where $\langle r^l \rangle$ is the 2nd moment of the $4f$ electron radial distribution and $\theta_l$ are geometrical factors tabulated for rare-earth ions~\cite{Taylorsi}.
Using the \textsc{mcphase} package~\cite{McPhasesi} we calculated the energy level scheme as well as the transition probabilities of ground-state CEF transitions for Yb$^{3+}$ in YbFeO$_3$ (Table~\ref{tab:CEF}).

Knowing the CEF parameters, one can calculate macroscopic properties of the Yb ions in a single-ion approximation~\cite{bauer2010magnetismsi}. We performed such calculations and found that the ground state doublet has a strong Ising-like anisotropy, with its easy-axis lying in the $ab$ plane with $\alpha=\pm21.1(6)^{\circ}$ to the $a$ axis.

 \begin{table}
 \caption{~The energy levels and out of ground state transition probabilities calculated  using the $B_l^m$ parameters, see text.} \label{tab:CEF}
 \begin{ruledtabular}
 \begin{tabular}{llll}
 & $|0 \rightarrow 1\rangle$ & $|0 \rightarrow 2\rangle$ & $|0 \rightarrow 3\rangle$  \\
\hline
$E$~(meV) & 17.2 & 53.3 & 64.7 \\
$\langle{}n| J_\perp |m\rangle^2$ & 3.09 & 0.28 & 0.08 \\
 \end{tabular}
 \end{ruledtabular}
 \end{table}

\subsection{``Shadow mode'' in the low-energy INS spectrum}

\begin{figure}[t!]
\includegraphics[width=1.0\columnwidth]{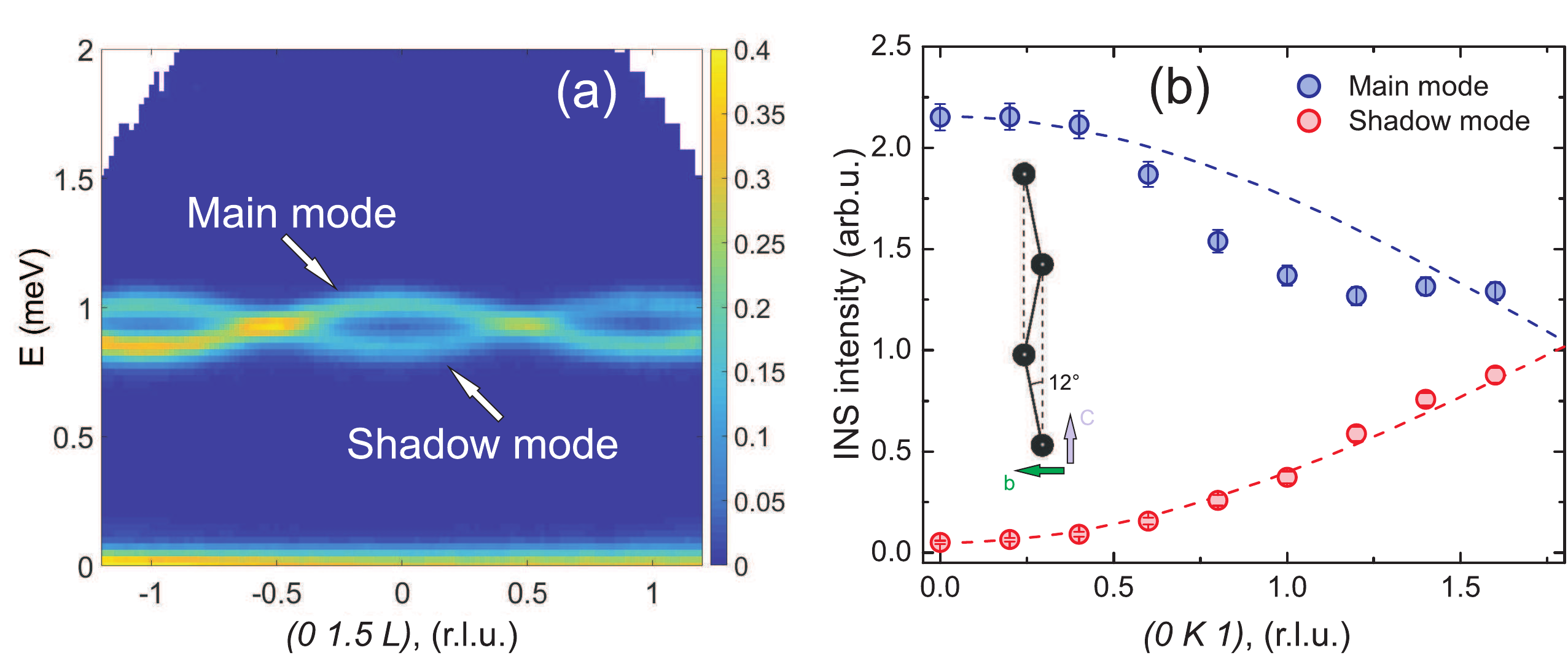}
  \caption{~``Shadow mode'' in the low-energy INS spectrum. (a) Experimental spectrum of YbFeO$_3$ at $T=2$~K along the $(0~1.5~L)$ direction. (b) Intensities of the ``main'' and ``shadow'' modes taken at $(0~K~1)$ as a function of $K$ (dots), and results of modeling with LSWT (solid lines)).}
  \label{ShadowMode}
\end{figure}

In this section we will discuss the appearance of a second ``shadow'' mode in the spectrum of Yb excitations. The low-energy INS spectrum taken at $T=2$~K along the $(0~1.5~L)$ direction is shown in Fig.~\ref{ShadowMode} (left) and Fig.~\ref{ShadowMode2}.
One can see that in addition to the ``main'' mode, which is clearly visible for both zero and non-zero $K$ [Fig.~3(a) in the main text and Fig.~\ref{ShadowMode}(a), respectively], there is a second mode with a similar dispersion and periodicity, which is shifted by wave-vector ${L}~\rightarrow~L+1$.
Figure~\ref{ShadowMode}(b) shows the intensities of the INS peaks at $\mathbf{Q}=(0~K~1)$, which corresponds to the ``main'' and ``shadow'' modes as function of $K$, and one can see that the increasing of $K$ continuously suppresses the intensity of the ``main'' mode, while the second mode appears at non-zero $K$ and its intensity is increasing with $K$. A similar phenomenon was previously observed in CoNb$_2$O$_6$~\cite{Cabrerasi}. The authors associated the emergence of the second ``shadow'' mode with the buckling of magnetic chains, where consecutive ions along the chain are alternatively displaced by $\pm~\zeta{}b$. Such a zig-zag structure in the $b$ direction leads to a doubling of the magnetic unit cell along the $c$ axis and appearance of a second mode, which has non-zero intensity only for finite $K$.

Yb$^{3+}$ ions in YbFeO$_3$ are also buckled by $\pm$0.378~\AA\ along the $b$ axis and form zig-zag structure [see insert in Fig.~\ref{ShadowMode}(b)]~\cite{marezio1970crystalsi}.
We model the $\mathbf{Q}$-dependence of intensities of both modes with LSWT using \textsc{SpinW} software~\cite{Tothsi}, taking into account a single intrachain exchange interaction and molecular field of the Fe subsystem ($J_z=0.25$~meV, $J_{xy}=0.1$~mev, $H=1.2$~meV). Experimental intensities and the results of LSWT calculations are in perfect agreement, as seen in Figs.~\ref{ShadowMode}(b) and Fig.~\ref{ShadowMode2}, proving that the low-energy dispersion along the $(00L)$ direction could be clearly associated with the nearest neighbor Yb-Yb exchange interaction along the $c$ axis.

\begin{figure}[b]\vspace{-0.3em}
\noindent\animategraphics[controls, loop, width=\columnwidth]{1}{Shad_}{1}{9}
\vspace{-0.3em}
\caption{~Animation illustrating observed (left) and calculated (right) excitation spectra of YbFeO$_3$ at $T=1.7$~K.  Energy slice along the $(00L)$ direction with $(H00)$ integrated over the range [$-0.25$, 0.25] and $(0K0)$ over the range [$K-0.25$, $K+0.25$], $K$ takes values $K=$0, 0.25,..2.  }
\label{ShadowMode2}
\end{figure}

\subsection{Interchain coupling}

\begin{figure}[tb]
    \includegraphics[width=0.9\columnwidth]{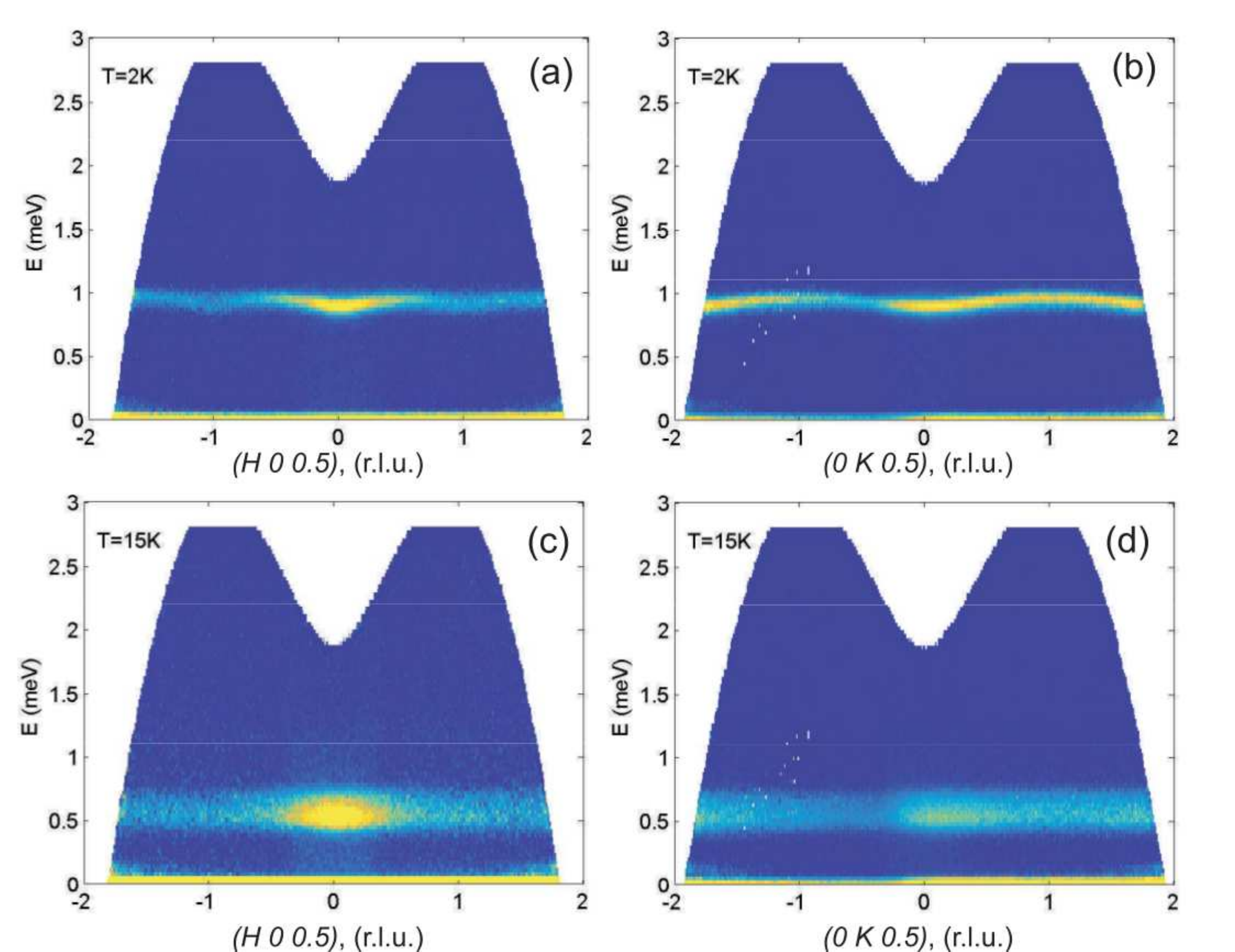}
    \caption{~Observed excitation spectra of YbFeO$_3$ at $T=2$~K (top) and $T=10$~K (bottom) along $(H0\frac{1}{2})$ (left) and $(0K\frac{1}{2})$ (right) directions.  $K, L$ (in left)  and $H, L$ (in right) were integrated over the range [-0.1, 0.1] and [0.4, 0.6] (r.l.u.), respectively.
    }
    \label{si_hk0}
\end{figure}

The INS intensity along $H$ and $K$ directions for $L=\frac{1}{2}$ are shown in Fig.~\ref{si_hk0}. For decoupled magnetic chains one should expect non-dispersive excitations, and the data show only a weak dispersion in both directions, indicating that interchain coupling is indeed small.

\subsection{Magnetization data}

\begin{figure}[b]
    \includegraphics[width=1\columnwidth]{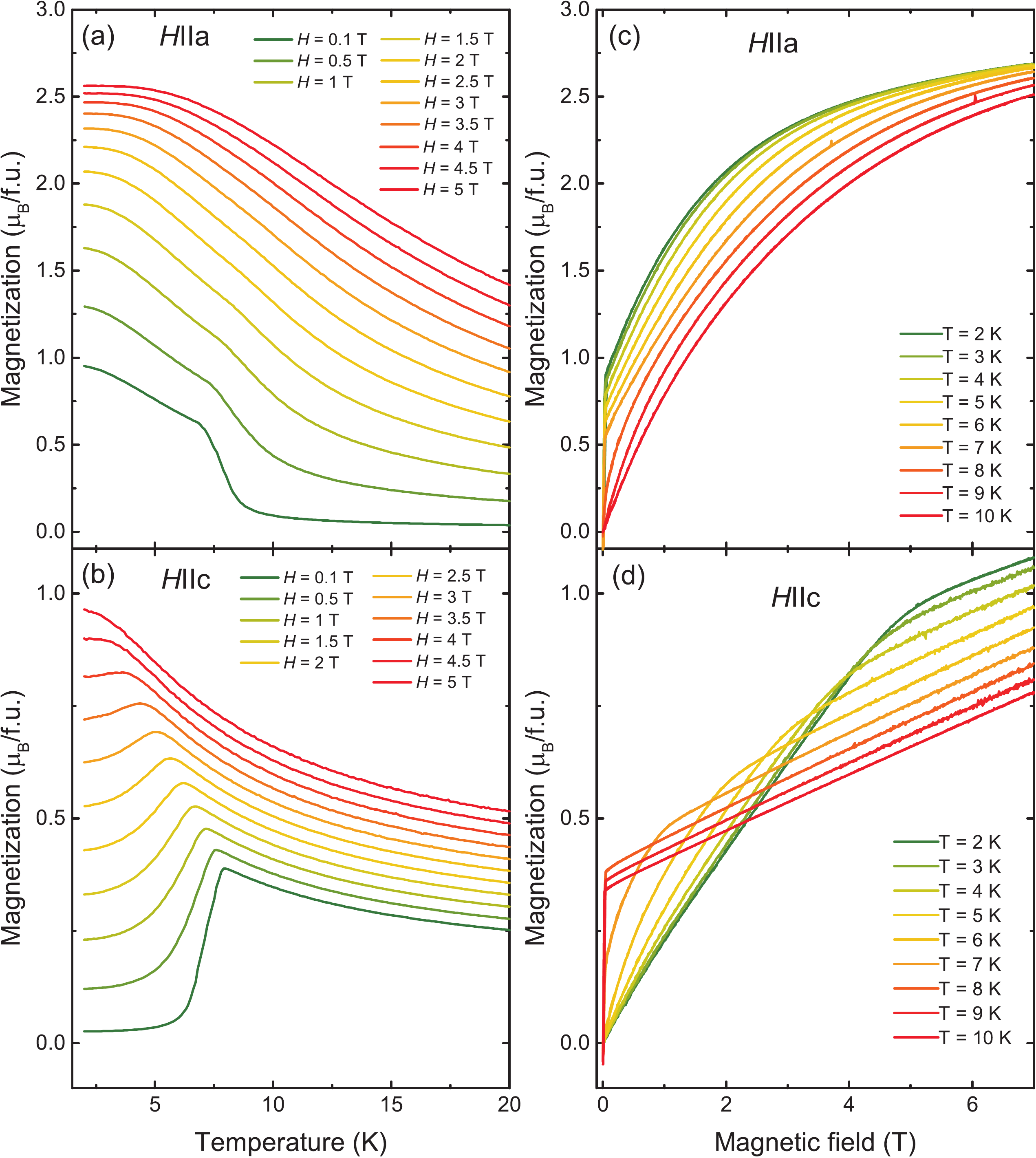}
    \caption{~Magnetization of YbFeO$_3$ measured as a function of temperature and magnetic field along the $a$ and $c$ axes.}
    \label{Magnetization}
\end{figure}

Figure~\ref{Magnetization} presents a summary of the magnetization measurements of YbFeO$_3$.
Temperature dependences of magnetization along $a$ and $c$ axes are shown in Fig.~\ref{Magnetization}(a,b).
At low fields, decreasing temperature induces the SR transition, which is seen as a sudden drop and raise of magnetization for the $\mathbf{H}\parallel{}[100]$ and $\mathbf{H}\parallel{}[001]$, respectively.
Increasing field smeared out these features.
An application of the magnetic field along the $c$ axis stabilizes the $\rm \Gamma4$ phase and gradually decreases $T_{\mathrm{SR}}$.
We followed critical points on the temperature and magnetic field dependences of the magnetization [Fig.~\ref{Magnetization}(b,d)] and reconstructed a phase diagram of YbFeO$_3$, as shown in Fig.~6(a) in the main text.

Magnetic field dependencies of the magnetization in the $\mathbf{H}\parallel{}[100]$ and $\mathbf{H}\parallel{}[001]$ directions are shown in Fig.~\ref{Magnetization}(c,d).
At $T>T_{\mathrm{SR}}$ and $\mathbf{H}\parallel{}[001]$ the magnetization has a weak low-field net moment $M^c_{\mathrm{FM}} \approx 0.3~\mu_{\rm B}/{\mathrm f.u.}$ due to canting of the Fe magnetic moments, similar to other orthoferrites~\cite{bozorth1958magnetizationsi, Whitesi}.
Field application induces a linear increase of the magnetization.
Below the $T_{\mathrm{SR}}$, net moment is directed along the $a$ axis and application of the magnetic field along the $c$ axis induces the SR transition from $\rm \Gamma2$ to $\rm \Gamma4$ (seen as a kink in the $M(H)$ curves).
Above the SR transition the magnetization increases linearly, with a slope of the $\frac{dM}{dH} = 0.05~\mu_{\rm B}/{\mathrm f.u.\cdot{}T}$

Temperature and field dependences of the magnetization, measured at magnetic fields applied along the $a$ axis are shown in Fig.~\ref{Magnetization}(a,c).
Below $T_{\mathrm{SR}}$ the net moment of $M^a_{\mathrm{FM}} \approx 1\mu_{\rm B}/{\mathrm f.u.}$ is about three times larger compared to $M^c_{\mathrm{FM}}$.
Increasing field induces magnetization with a Brillouin-like shape and saturation moment of $M_{\rm S} \approx 2.75\mu_{\rm B}/{\mathrm f.u.}$

\subsection{Specific heat measurements}
\begin{figure}[tb]
    \includegraphics[width=1.0\columnwidth]{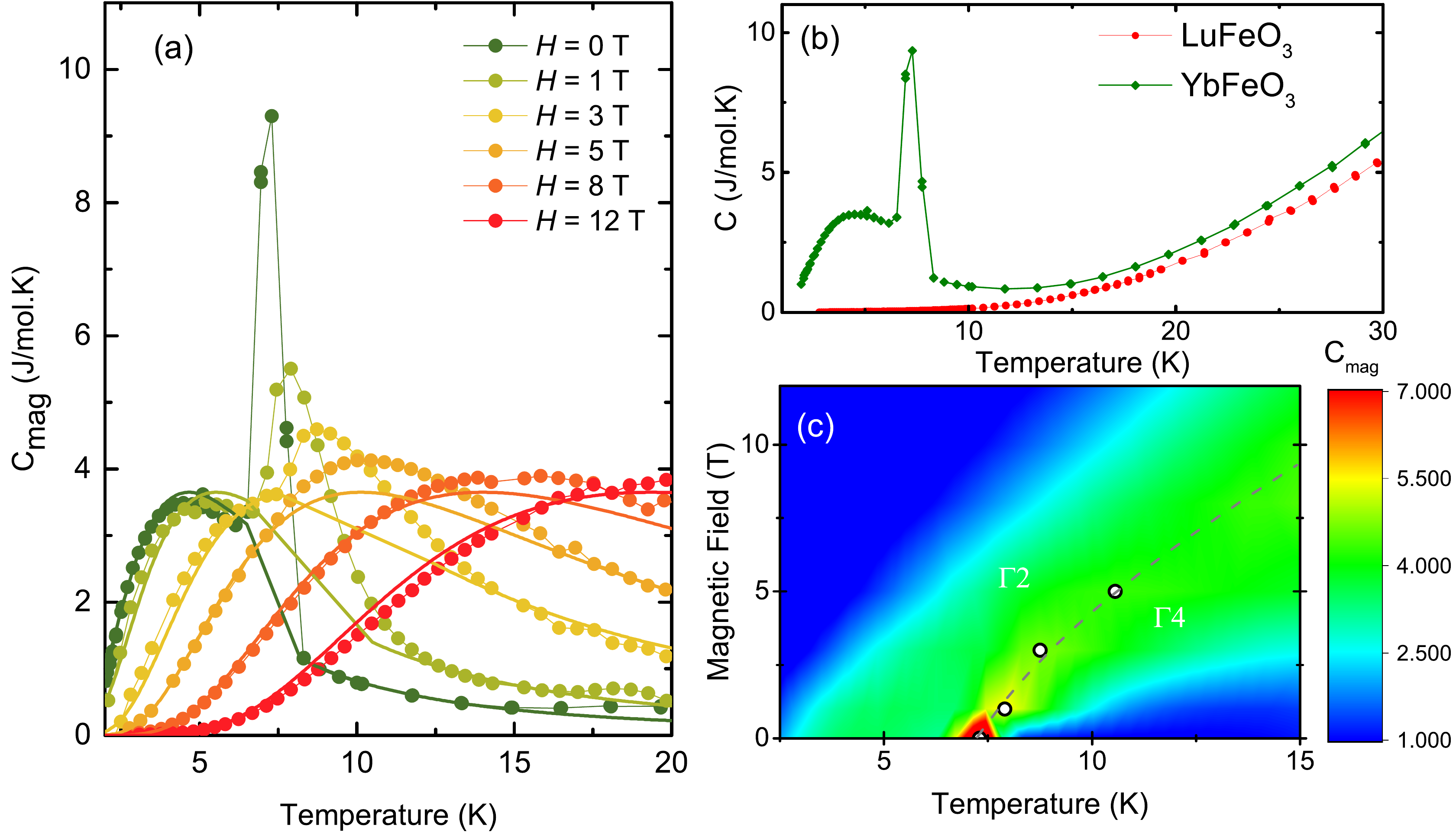}\vspace{3pt}
    \caption{~Results of the specific heat measurements of YbFeO$_3$ with magnetic field applied along the $a$ axis.
    (a) Temperature dependencies of the magnetic contribution to the specific heat. Solid curves show specific heat calculated for the two-level system (see main text).
    (b) Low-temperature specific-heat curves of YbFeO$_3$ and LuFeO$_3$.
    (c) $H-T$ phase diagram of YbFeO$_3$. The colorplot shows specific heat data.
    }
    \label{Specific_heat}
\end{figure}

The specific heat of YbFeO$_3$  is the sum of three different contributions that vary with temperature: the lattice contribution, the magnetic specific heat of the iron subsystem and the specific heat of the rare-earth subsystem.
In order to exclude the first two terms, we measured the specific-heat of the isostructural compound LuFeO$_3$ as a reference with a non-magnetic $R^{3+}$ ion [see Fig.~\ref{Specific_heat}(b)].
A magnetic signal, associated with the specific heat of the Yb subsystem and a sharp $\lambda$-peak, caused by the SR transition, is shown in Fig.~\ref{Specific_heat}(a) for different values of external field, applied along the $a$ axis.
The modification of the $\lambda$-peak in magnetic field was studied in detail previously~\cite{Moldoversi}.
Here, following the position of the $\lambda$-peak, we reconstructed a magnetic-field--temperature phase diagram of YbFeO$_3$  [see Fig.~\ref{Specific_heat}(c)] and calculated the effective values of the Yb ground state splitting as function of external field (see Table~\ref{Specific_heat_table}).

To calculate the specific heat of the Yb subsystem in the general case, one should perform numerical calculations based on the full magnetic Hamiltonian including CEF term, exchange interaction, magnetic field etc.
However, the Yb subsystem in YbFeO$_3$ is strongly polarized by a molecular field of the Fe subsystem.
The Zeeman splitting of the ground-state doublet exceeds the Yb-Yb exchange energy, since the gap size is much larger than the magnon band width.
Furthermore, due to the large CEF splitting $\Delta E=20$~meV, only the low-lying doublet gives a contribution to the specific heat at low temperatures $T<20$~K.
In this case, we can simplify the model and use a two-level Schottky anomaly expression with a single parameter $\Delta$ to describe the low-$T$ specific heat
\begin{eqnarray}
C = {R} \Big(\frac{\Delta}{T}\Big)^2 \frac{\exp{\{\Delta/T\}}}{[1 + \exp{\{\Delta/T\}}]^2},
\label{Schottky}
\end{eqnarray}
where $R$ is the universal gas constant and $\Delta$ is a ground-state doublet splitting.
The fitted curves are shown by the solid lines in Fig.~\ref{Specific_heat}.
There is a good agreement between the experimental and calculated data, excluding the SR transition anomaly, which was not included into the calculation.
The $\Delta$ values for the different magnetic fields are shown in Table~\ref{Specific_heat_table}.
We found the effective $g$-factor for the $\rm \Gamma2$ phase $g_a^{\rm \Gamma2}=4.39$, in a reasonable agreement with the results of the INS measurements.

\begin{table}\vspace{-2pt} 
\caption{~Energy gap $\Delta$ for the Yb ground-state doublet splitting, derived from specific-heat measurements.}
\label{Specific_heat_table}
\begin{ruledtabular}
\begin{tabular}{ccc}
$H$(T)    &  $\Delta$~(meV),  $T<T_{\mathrm{SR}}$     & $\Delta$~(meV), $T>T_{\mathrm{SR}}$ \\
\hline
 ~0        &          0.97  &             0.57 \\
 ~1        &          1.15  &             0.82 \\
 ~3        &          1.65  &             1.50 \\
 ~5        &          2.09  &             2.09 \\
 ~8        &          2.90  &             2.90 \\
 12        &          4.00  &             4.00 \\

\end{tabular}
\end{ruledtabular}
\end{table}

\subsection{Excitation spectra of the 1D-XXZ Hamiltonian in the field-polarized state}

\begin{figure}[tbh!]
    \includegraphics[width=1\columnwidth]{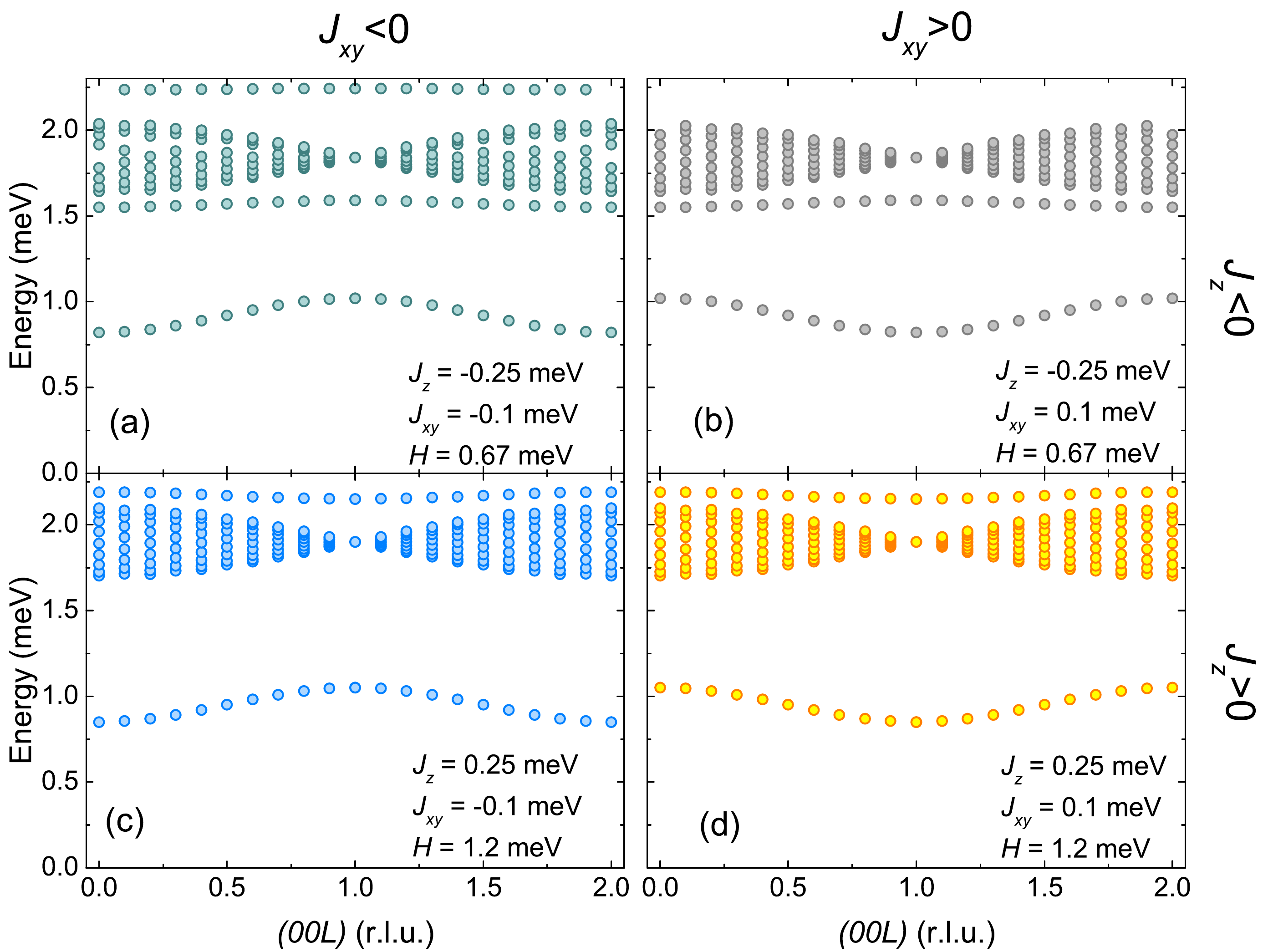}
    \caption{~Eignenvalues of the 1D-XXZ Hamiltonian calculated for the 4 sets of parameters. Values of $J_z$, $J_{xy}$ and $H$ are indicated in the figure.
    }
    \label{ALPS}
\end{figure}

We fitted the INS data to the model spectra and calculated a set of exchange parameters.
In order to describe the excitation spectrum of the Yb subsystem we use a 1D XXZ spin Hamiltonian (see Eq.~(4) in the main text).
As we mentioned above, at temperatures $T<T_{\mathrm{SR}}$, the magnetic field $\mathbf{H}$ is longitudinal to the easy axis of the Yb moments and exceeds $J_z$.
Accordingly, the ground state of the Yb spin chain is a field-polarized FM state.
Excitation spectra of the FM XXZ spin chains in magnetic fields have been studied in detail previously~\cite{Orbachsi,Schneidersi, schneider1982excitationsi, Torrance587si}.

In this section we present finite-chain exact diagonalization calculations of the eigenstates of Eq.~7 (main text), using the Lanczos algorithm realized in \textsc{alps} software~\cite{ALPS1si, ALPS2si}.
We investigated a chain with $L=20$ sites with three parameters: $J_z$, $J_{xy}$ and $\mathbf{H}$, longitudinal to the $z$-direction.
We fitted this set of parameters in order to qualitatively describe the positions of the single magnon mode and the two-magnon continua.

There are three types of excitations in the experimental INS data: i) an intense single-particle mode with the maximum gap at the $\Gamma$-point; ii) a dispersionless excitation at $E=1.6$~meV, which we associate with the two-magnon bound state~\cite{schneider1982excitationsi} and iii) a broad continuum in the energy range $E\approx1.7-2.1$~meV.

Figure~\ref{ALPS} shows calculated spectra for the four different sets of parameters of the Hamiltonian~(Eq.9 in the main text) (All possible combinations $J_z\gtrless0$, $J_{xy}\gtrless0$).
Figure~\ref{ALPS}(a) shows calculated spectra for $J_z, J_{xy} < 0$, which corresponds to the classical FM XXZ chain~\cite{Schneidersi,schneider1982excitationsi}.
In this case, the single particle mode has a minimum at the $\Gamma$-point, in contrast to our experimental results.
Changing $J_{xy}$ to AFM, we got the best fit of our spectrum as clearly seen in Fig.~\ref{ALPS}(b).
When both $J_z$ and $J_{xy}$ are AFM, we still obtain the correct dispersion for the single particle mode, whereas the dispersionless excitation is shifted to a higher energy above the continuum in contradiction with the experiment, Fig.~\ref{ALPS}(d).

For $T>T_{\mathrm{SR}}$, when the molecular field is transverse to the Yb easy axis, we were not able to reproduce the details of the excitation spectra even qualitatively.
The excitation energy $E\approx7$~K is comparable with the temperature of the measurements.
Apparently, our zero-temperature calculations failed to reproduce the experimental data.

\subsection{Free energy functional at $T$ close to $T_\mathrm{SR}$}

\begin{figure}[tbh!]
    \includegraphics[width=1\columnwidth]{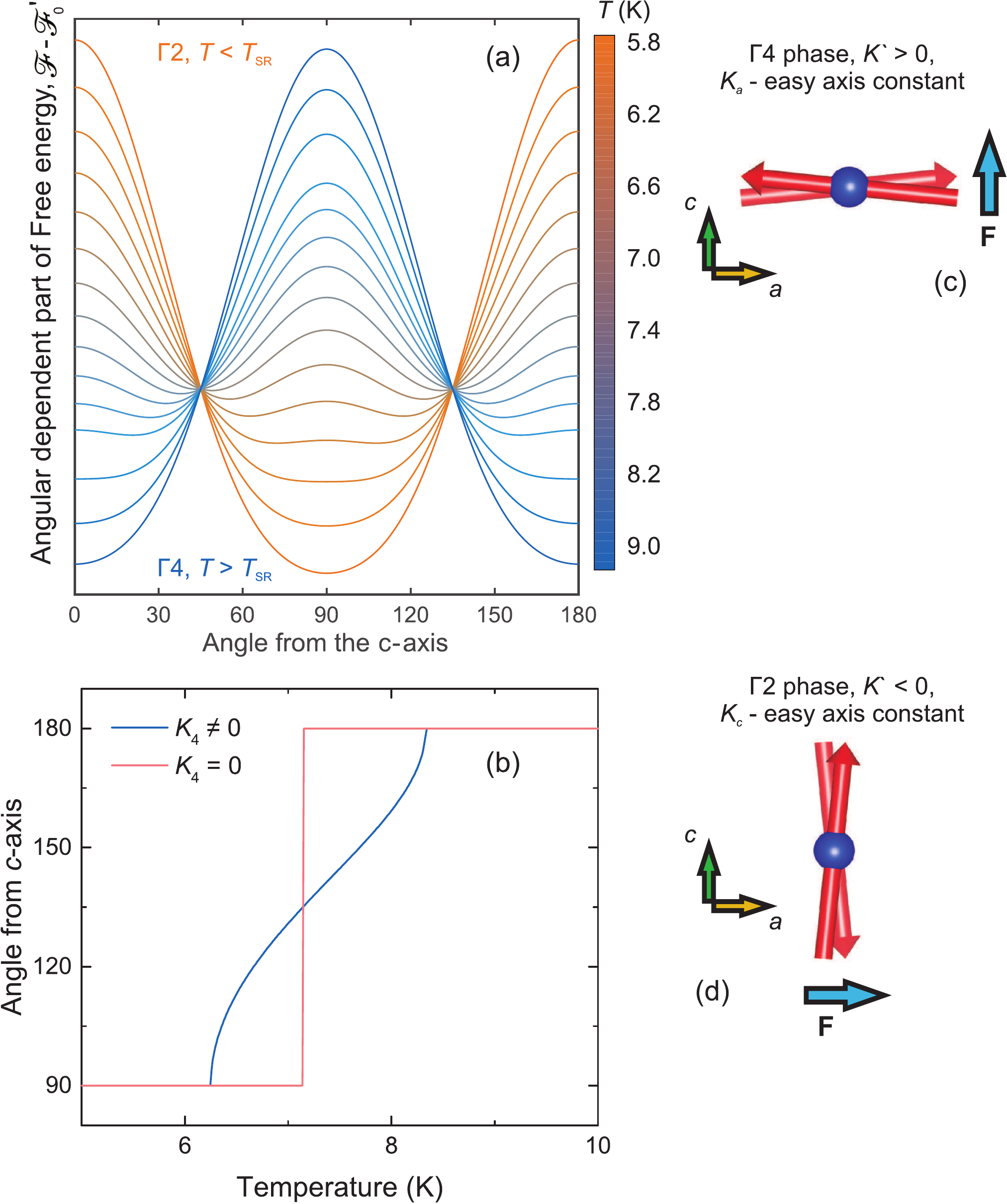}
    \caption{
    (a)~Free energy $\mathcal{F}$-$\mathcal{F}_0$ [Eq.~(\ref{FreeEnergy1})] as a function of angle between $c$axis and net moment $\mathbf{F}$ plotted at temperatures close to $T_{\mathrm{SR}}$.
    (b)~Angle between $c$ axis and net moment $\mathbf{F}$ as function of temperature, derived from the minimization of the Eq.~(\ref{FreeEnergy1}) for $K_4~=~0$ (red) and $K_4~>~0$ (blue).
    (c,~d)~Sketch of the magnetic structure of the Fe subsystem at temperatures above and below $T_{\mathrm{SR}}$.
    }
    \label{Free_E}
\end{figure}

In this section we show how the $R$-Fe interaction could be redefined, assuming that the effective anisotropy constant of the Fe moments $K^{\prime}$ has a temperature dependence.
Recently, Bazaliy $et al$, proposed a modified mean-field theory in order to describe the SR transition in several rare-earth orthoferrites ($R$ = Tm Er, Yb)~\cite{bazaliy2004spinsi, bazaliy2005measurementssi, tsymbal2007magneticsi}.
They assumed that the weak net FM moment of the Fe subsystem $\mathbf{F}$ polarizes the paramagnetic $R$-subsystem, which, in turn, has a significant anisotropy of magnetic susceptibility.
Following this approach, we write down a free energy functional for YbFeO$_3$ at temperatures close to the SR transition in a form:
\begin{eqnarray}
 \mathcal{F} = && \mathcal{F}_0 - \frac{1}{2}(K_{a}-K_{c})\mathrm{cos}(2\theta) - \frac{1}{2}(K_{4})\mathrm{cos}(4\theta) \nonumber\\
  + && \frac{\beta}{2}(\chi^2_cF^2_c+\chi^2_aF^2_a),
  \label{FreeEnergy1}
\end{eqnarray}
where $\mathcal{F}_0$ is angular independent part, $\theta$ is an angle between the moment $\mathbf{F}$ and the $c$ axis, $K_c$ and $K_a$ are single-ion anisotropy constants of Fe ions, $\beta$ -- effective field, which couples Fe and Yb moments, and $\chi\propto\frac{1}{T}$ is the temperature dependent anisotropic magnetic susceptibility tensor of the Yb moments.
The fourth-order anisotropy constant $K_{4}$ is a few orders of magnitude smaller than the second-order $K_a$ and $K_c$, and plays a crucial role only at a temperature, where the second order term $\propto \mathrm{cos}(2\theta)$ converges to zero and the SR transition takes place~\cite{bazaliy2004spinsi, Belovsi}.
Assuming that the moment $|\mathbf{F}|$ is conserved through the SR transition~\cite{bazaliy2004spinsi}, one could rewrite $F_a~=~|\mathbf{F}|\mathrm{sin}(\theta)$ and $F_c~=~|\mathbf{F}|\mathrm{cos}(\theta)$.
With these approximations, Eq.~(\ref{FreeEnergy1}) could be rewritten in a simpler form:
\begin{eqnarray}
\mathcal{F} = \mathcal{F}_0^{\prime} - K^{\prime}\mathrm{cos}(2\theta)  - \frac{1}{2}(K_{4})\mathrm{cos}(4\theta),
\label{FreeEnergy2}
\end{eqnarray}
where the new angular independent term $F_0^{\prime}$ and the effective anisotropy constant $K^{\prime}$ are defined as:
\begin{eqnarray}
\mathcal{F}_0^{\prime} = \mathcal{F}_0 + \frac{\beta{}|\mathbf{F}|^2}{4}(\chi^2_c + \chi^2_a),
\label{F_new}
\end{eqnarray}
\begin{eqnarray}
K^{\prime} =  \frac{1}{2}(K_{a} - K_{c} - \frac{\beta |\mathbf{F}|^2}{4}(\chi^2_a - \chi^2_c)).
\label{K_new}
\end{eqnarray}

It is known that $K_a>K_c$ ($\chi=0$) for the orthoferrites with nonmagnetic $R$.
Therefore $K^{\prime}>0$ tends to order the Fe moment along the $a$ axis and stabilizes the $\rm \Gamma4$ magnetic phase.
This situation is schematically shown in Fig.~\ref{Free_E}(c).
On the other hand, in YbFeO$_3$ $\chi\neq0$ and the CEF leads to a strong anisotropy of the magnetic susceptibility $\chi_a~\gg~\chi_c$.
The anisotropy can be seen from the results of magnetic measurements (see Fig.~\ref{Magnetization}), CEF calculations (see section~\ref{sec_CEF}), as well as from the polarization factor of neutron scattering (see Fig.~4 in the main text).
At high temperatures, the magnetic susceptibility of Yb is small, and $K_a$ dominates in $K^{\prime}$, similar to orthoferrites with nonmagnetic $R$.
Upon cooling, the magnetic susceptibility of the Yb moments is increasing and this, in turn, would decrease the value of $K^{\prime}$ and the SR transition takes place when $K^{\prime}=0$.
Further temperature decreasing would change the sign of $K^{\prime}$, making an easy axis along the $c$ direction.
Angular dependences of the free energy and sketches of the magnetic structures are schematically shown in Fig.~\ref{Free_E} for both cases, $T>T_\mathrm{SR}$ and  $T<T_\mathrm{SR}$.
We found that the high-temperature phase $\rm \Gamma4$ is stabilized by the dominating effective $K_a$ constant, whereas below the $T_{\rm SR}$, type of anisotropy is changed, making the $c$ direction a new easy axis with domination of $K_c$.

Thus, there are four main interactions, which influence the magnetic properties of the Fe subsystem: the strong Heisenberg exchange interaction $J_\text{Fe-Fe}$, DM exchange $D$, single-ion anisotropy $K$ and Yb-Fe interaction $J_\text{Fe-Yb}$.
The energy hierarchy is $J_\text{Fe-Fe} \gg D > K \approx J_\text{Fe-Yb}$.
Note, that this simple ``mean-field''-like analysis has only two assumptions: i) strong anisotropy of the Yb susceptibility and ii) polarized by a net FM moment Yb subsystem.
It shows that we can take into account the influence of $J_\text{Fe-Yb}$ on the Fe subsystem via the effective renormalization of the anisotropy constant of the Fe subsystem.

\end{document}